\documentclass[iop,superscriptaddress]{emulateapj}
\usepackage{amsmath,epsfig,url,natbib}
\usepackage{graphicx,epstopdf,float,color}
\journalinfo{Accepted for publication in the Astrophysical Journal}
\submitted{}
\def\lsim{\mathrel{\rlap{\lower4pt\hbox{\hskip1pt$\sim$}}
    \raise1pt\hbox{$<$}}}                
\def\gsim{\mathrel{\rlap{\lower4pt\hbox{\hskip1pt$\sim$}}
    \raise1pt\hbox{$>$}}}                
\begin{document}
\title{A Study of Massive and Evolved Galaxies at High Redshift}
\author{H. Nayyeri\altaffilmark{1}, B. Mobasher\altaffilmark{1}, S. Hemmati\altaffilmark{1}, S. De Barros\altaffilmark{1}, H. C. Ferguson\altaffilmark{2},  T. Wiklind\altaffilmark{2}, T. Dahlen\altaffilmark{2}, M. Dickinson\altaffilmark{3}, M. Giavalisco\altaffilmark{4}, A. Fontana\altaffilmark{5}, M. L. N. Ashby\altaffilmark{6}, G. Barro\altaffilmark{7}, Y. Guo\altaffilmark{7}, N.P. Hathi\altaffilmark{8}, S. Kassin\altaffilmark{2}, A. Koekemoer\altaffilmark{2}, S. Willner\altaffilmark{6}}

\altaffiltext{1}{University of California Riverside, Riverside, CA 92512}
\altaffiltext{2}{Space Telescope Science Institute, Baltimore, MD 21218}
\altaffiltext{3}{National Optical Astronomy Observatory, Tucson, AZ 85719}
\altaffiltext{4}{University of Massachusetts Amherst, Amherst, MA 01003}
\altaffiltext{5}{INAF Rome Observatory, Rome, Italy}
\altaffiltext{6}{Harvard-Smithsonian Center for Astrophysics, Cambridge, MA 02138}
\altaffiltext{7}{University of California Santa Cruz, Santa Cruz, CA 95064}
\altaffiltext{8}{Laboratoire d’Astrophysique de Marseille, Marseille, France}

\begin{abstract}

We use data taken as part of HST/WFC3 observations of the Cosmic Assembly Near-infrared Deep Extragalactic Legacy Survey (CANDELS) to identify massive and evolved galaxies at $3<z<4.5$. This is performed using the strength of the Balmer break feature at rest-frame 3648\AA, which is a diagnostic of the age of the stellar population in galaxies. Using WFC3 H-band selected catalog for the CANDELS GOODS-S field and deep multi-waveband photometry from optical (HST) to mid-infrared (Spitzer) wavelengths, we identify a population of old and evolved post-starburst galaxies based on the strength of their Balmer breaks (Balmer Break Galaxies- BBGs). The galaxies are also selected to be bright in rest-frame near-IR wavelengths and hence, massive. We identify a total of 16 BBGs. Fitting the spectral energy distribution (SED) of the BBGs show that the candidate galaxies have average estimated ages of $\sim$ 800 Myr and average stellar masses of $\sim 5 \times 10^{10} M_{\odot}$, consistent with being old and massive systems. Two of our BBG candidates are also identified by the criteria that is sensitive to star forming galaxies (LBG selection). We find a number density of $\sim 3.2 \times 10^{-5}\:Mpc^{-3}$ for the BBGs corresponding to a mass density of $\sim 2.0 \times 10^{6}\:M_{\odot}/Mpc^3$ in the redshift range covering the survey. Given the old age and the passive evolution, it is argued that some of these objects formed the bulk of their mass only a few hundred million years after the Big Bang.  

\end{abstract}
\maketitle

\section{Introduction}

Understanding the evolution of galaxies with look-back time is among the greatest challenges in observational astronomy today. According to the $\Lambda$-CDM scenario, galaxies assembled most of their mass through sequential merging, becoming more massive as they move down to lower redshifts \citep {Kauffmann1993, Reed2003}. These hierarchical models predict that dark matter halos become ever more massive as we move to the more recent past, with the gravitational potential wells produced by these halos providing the required seeds for gas collapse, leading to formation of galaxies \citep {Sheth2001, Navarro1996}. As a result, more massive galaxies are expected to form in massive halos at lower redshifts with younger and less massive systems often located at a more distant past (i.e., higher redshifts). However, recent studies have indicated a new population of massive galaxies at redshifts as high as $z \sim 5$ when the Universe was $\sim$ 1 Gyr old \citep {Mobasher2005, Wiklind2008, Caputi2012}. Furthermore, some of these galaxies show evidence for an evolved stellar population, formed through an initial burst of star formation followed by passive evolution \citep {Labbe2005, Mancini2009, Fontana2009, Guo2012, Stefanon2013, Straatman2014, Oesch2013, Huang2011}. This implies that most of these objects formed the bulk of their mass at very high redshifts.

A widely used method for selecting high redshift ($z > 3$) galaxies is based on the identification of Lyman break feature, the so-called dropout technique \citep {Steidel1996, Giavalisco2004b, Bouwens2007, Beckwith2006}. Many of the evolved systems at high redshifts lack the strong UV continuum characteristic of star forming galaxies and therefore will be missed by such surveys. These evolved systems could strongly constrain galaxy formation scenarios and may constitute a significant fraction of the mass density of galaxies \citep {Ilbert2013}. Studying the contribution of this population to the global number and mass density will strongly constrain the high-mass end of the galaxy mass function at high redshifts. If such evolved and massive systems, missed from high redshift Lyman break selected surveys, exist in large numbers, they could put strong constraints on evolutionary scenarios and predictions from hierarchical models for galaxy formation.
 
We propose to identify these evolved systems by searching for the pronounced Balmer break feature at rest-frame 3648\AA\, which is an age dependent diagnostic of their stellar population (hence the name Balmer Break Galaxies; BBGs). The Balmer break is most prominent in A-type stars with surface temperatures $\sim 10000^{\circ}$K. The strength of the Balmer break feature does not monotonically increase with age and it reaches a maximum in stellar populations with ages 0.3 to 1 Gyr. Therefore it is an efficient tool for identifying relatively quiescent (quenching) systems with current star formation rates significantly lower than the past average \citep {Kriek2008} . 

In this work, we use the new Hubble Space Telescope/Wide Field Camera 3 (WFC3) near infrared data from observations taken as a part of the Cosmic Assembly Near-infrared Deep Extragalactic Legacy Survey \setcounter{footnote}{0}\footnote{http://candels.ucolick.org} (CANDELS: PI. S. Faber and H. Ferguson; see \citealp {Grogin2011} and \citealp {Koekemoer2011}) program to identify redshifted Balmer break features at $3.0<z<4.5$. In this redshift range the Balmer break falls between the H and K-band filters. Here we also take advantage of the very deep $K_s$-band observations (at $\sim$ 2.2 $\mu$m) red-ward of the Balmer break from the VLT/HAWK-I instrument in the GOODS-S (Fontana et al. in prep.). The very deep near-IR photometry (from WFC3 and HAWK-I) ensures an accurate measurement of the Balmer break amplitude. Also the HST/WFC3 and optical data (blue-ward of the break) help to break the degeneracy and avoid contamination by lower redshift dusty galaxies. This redshift range was chosen to take full advantage of the available deep WFC3 data when selecting the BBG candidates. This also reduces the lower redshift red color contaminants in the sample. Finally, since the objects are brighter at these lower redshifts, it provides a higher likelihood of successful spectroscopic observation compared to higher redshift (fainter) candidates \citep {Wiklind2008}.

The search for massive galaxies at high redshift is greatly aided by the availability of deep infrared data from the Spitzer Space Telescope. Massive galaxies are brightest in their rest-frame near-IR, which is shifted to mid-infrared Spitzer bands at higher redshifts. In this respect, Spitzer IRAC \citep {Fazio2004} provide excellent probes of the stellar mass of galaxies at high redshifts \citep {Papovich2006}. Here we use Spitzer images, specifically the very deep IRAC observations at 3.6$\mu$m and 4.5$\mu$m, which is essential for estimating the stellar mass of these candidates.

This paper is organized as follows. In Section 2 we discuss the photometric data used for this study. In Section 3, we develop the criteria used to select the BBG candidates and discuss the robustness of the selection criteria. Section 4 presents the SED fitting of the candidate BBGs identified from the color-color plots and discuss the candidate properties. In Section 5 we discuss the MIPS detected BBG sample. We compare properties of the galaxies selected through the LBG and BBG techniques over the same redshift range in section 6 and we discuss the results in Section 7. Throughout this paper we assume a standard cosmology with $H_0=70\:kms^{-1}Mpc^{-1}$, $\Omega_m=0.3$ and $\Omega_\Lambda=0.7$. All magnitudes are in the AB system where $m_{AB}=23.9-2.5log(f_{\nu}/1\mu Jy)$ \citep {Oke1983}.

\section{The Photometric Data}

We use the multi-waveband deep photometric observations in the GOODS-S field \citep {Giavalisco2004a}. In this study we focus on the CANDELS GOODS-S Deep area, which has very deep observations by the VLT HAWK-I instrument (Fontana et al. in prep.). The WFC3/IR observations in the CANDELS GOODS-S deep area consist of a rectangular grid of $3 \times 5$ tiles ($\sim 6.8^{\prime} \times 10^{\prime}$; \citealp {Grogin2011}) and photometric observations in more than fourteen broadband filters. In the redshift range of interest here, we explore a volume of $\sim 2.05 \times 10^5\: Mpc^3$. In the following we describe details of the data and the construction of the photometric catalog. 

\subsection{U-band}

The U-band data in GOODS-S were taken using the VIMOS instrument on the Very Large Telescope \citep {Nonino2009}. They reach a magnitude limit of U=27.97 (5$\sigma$ over 0.8 arcsec radius aperture). These data are much deeper than the previous CTIO $U$-band observations available in the GOODS-S ($U \sim 26.53$), and have a narrower PSF. 

\subsection{Optical Data}

The optical data in GOODS-S are from observations by the HST/ACS instrument. The observations are done in the F435W ($B_{435}$), F606W ($V_{606}$), F775W ($i_{775}$) and F850LP ($z_{850}$) filters to magnitude limits of 28.95, 29.35, 28.55 and 28.55 mag respectively (with 5$\sigma$ detections in FWHM sized apertures; \citealp {Guo2013}). We use version 3.0 ACS images, which are the deepest in the optical bands.

\subsection{Near and Mid  Infra-Red Data}

Near infrared data for GOODS-S were taken with the Wide Field Camera 3 infrared instrument on the Hubble Space Telescope as part of the CANDELS treasury program \citep {Koekemoer2011, Grogin2011}. The observations are in the F105W ($Y_{105}$), F125W ($J_{125}$) and F160W ($H_{160}$) filters. The 5$\sigma$ limiting magnitudes are 28.45 in $Y_{105}$, 28.34 in $J_{125}$ and 28.16 in $H_{160}$ \citep {Guo2013}. These are the deepest near-IR data in the GOODS-S field, being more than 2 magnitudes deeper than previous data from the VLT/ISAAC in the J and H bands. 

The GOODS-S has also been observed in the near-IR by the VLT HAWK-I in the $K_s$ filter at effective wavelength 2.2 $\mu$m (Fontana et al. in prep.). The HAWK-I $K_s$ magnitude limit is 26.45 over an aperture of 0.4 arcseconds \citep {Guo2013}. This is more than a magnitude deeper than the previous VLT/ISAAC $K_s$ data. The final field of view of this work is chosen to take advantage of the deep $K_s$ band observations by the VLT/HAWK-I. The depth variation of the HAWK-I across GOODS-S does not play a significant role in selecting our candidates because of the small variations in the sky standard deviation at the position of our sources.

The mid infrared data for this study are from observations by the Spitzer Space Telescope IRAC instrument \citep {Fazio2004} of the GOODS fields covering the four IRAC bands (3.6 $\mu$m, 4.5 $\mu$m, 5.8 $\mu$m and 8.0 $\mu$m bands). Particularly important for this study are the IRAC 3.6 $\mu$m and 4.5 $\mu$m bands which are deeper given the data acquired during the Spitzer warm mission as part of the SEDS program \citep {Ashby2013} with 3$\sigma$ limiting magnitudes of 26 in both bands. The IRAC 5.8 $\mu$m and 8.0 $\mu$m has limiting magnitudes of 23.75 and 23.72 respectively \citep {Guo2013}. 

\subsection{The Photometric Catalog}

We use the WFC3 $H$-band selected catalog for GOODS-S field which is generated using the Template-FITting (TFIT) algorithm \citep {Guo2013, Laidler2007}. TFIT is a robust method for combining multi wave-band data observed with different instruments. One of the complications of performing consistent multi-waveband photometry of objects is the difference in resolution and PSFs among different instruments. Objects that are well separated in high resolution images (ACS/WFC3) could be blended in the lower resolution data (ground based or IRAC). 

TFIT uses prior information from high resolution images to construct object templates that are then fitted to the lower resolution image from which fluxes are extracted \citep {Laidler2007}. Here we used the $H$-band selected catalog because in the redshift range of interest here, the WFC3 $H_{160}$ falls blue-ward of the Balmer break feature and hence, we need detections in this filter to be able to identify the BBGs. The final photometric catalog has data in more than 12 bands including VLT/VIMOS U, HST/ACS $B_{435}$, $V_{606}$, $i_{775}$ and $z_{850}$, HST/WFC3 $Y_{105}$, $J_{125}$ and $H_{160}$, VLT/ISAAC and HAWK-I $K_s$ and Spitzer/IRAC 3.6 $\mu$m, 4.5 $\mu$m, 5.8 $\mu$m and 8.0 $\mu$m bands \citep {Guo2013}.

\section{BBG Selection Criteria}

We use the strength of Balmer break feature, shifted to near infrared wavelengths, to develop a color criteria to identify post-starburst galaxies at $3.0<z<4.5$. The near infrared colors can also get contribution from redshifted emission lines to broadband filters. Although we do not generally expect to have a strong emission spectrum for passively evolving galaxies, these emissions can mimic the Balmer break feature, leading to false identification of the BBGs. The post starburst galaxy SEDs in the redshift range of interest could get contributions from several nebular emission lines, with the most prominent being [OII], [OIII] and H$\beta$. In this Section we develop the method and the criteria for selection of the BBG candidates and discuss the effect of nebular emission on the BBG templates. 

\subsection{Color-Color selection method}

The Balmer break feature in post-starburst galaxies causes a break in the broadband colors of galaxies that can be used to identify the BBGs. In the redshift range of this study, the Balmer break (at rest-frame 3648\AA) falls between the WFC3 $H_{160}$ and the HAWK-I $K_s$ filters. Therefore, we use ($H_{160}-K_s$), straddling the Balmer break feature, as the primary color for identifying the quiescing galaxies. Figure 1 shows a sample post-starburst galaxy template at redshift $z=3.9$ with relatively old stellar population (Age $\textgreater$ 350 Myr) and no dust extinction. The blue and red curves show the filter response functions of the WFC3 $H_{160}$ and the HAWK-I $K_s$ respectively. At the redshift of this galaxy these filters probe the blue and red sides of the Balmer break.

\begin{figure}
\begin{center}
\includegraphics[scale=0.4]{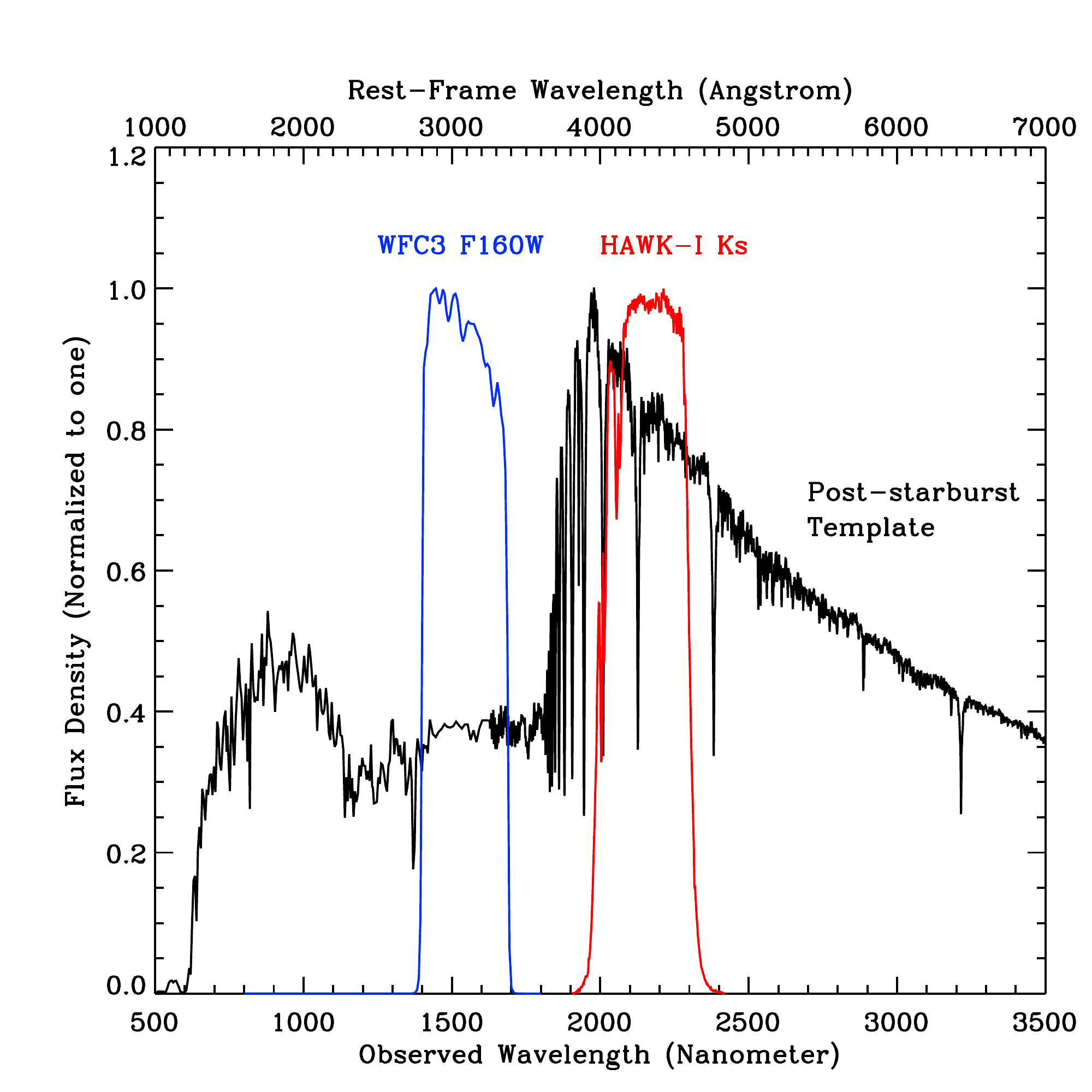}
\caption{A post-starburst galaxy template SED ($F_{\lambda}$) at $z=3.9$ (black) with the WFC3 $H_{160}$ (blue) and HAWK-I $K_s$ filters (red) straddling the Balmer break at observed 1.8 $\mu m$ (the plots have been normalized to a peak of one).}
\end{center}
\end{figure}

We use the synthetic spectral energy distribution code from Bruzual and Charlot 2003 (BC03; \citealp {Bruzual2003}) to create a library of model SEDs of galaxies spanning a wide range in parameter space and different galaxy types. We consider an exponentially declining star formation history (SFR $\propto exp(-t/\tau)$) for all of our models with the characteristic time scale for star formation, $\tau$, ranging from 0 to 200 Myr with steps of 25 Myr. We use a Salpeter initial mass function and use solar metallicity. We used the Calzetti attenuation curve \citep {Calzetti2000} for treatment of the extinction with color excess E(B-V) ranging from 0 to 0.9 (corresponding to $A_v=0-3.6$) in steps of 0.05. Attenuation due to neutral Hydrogen in the intergalactic medium was estimated using Madau's prescription \citep {Madau1995}. The age parameter, defined as the time since the onset of recent star formation, ranges from 0 to 2.5 Gyr (from 5 to 30 Myr in steps of 5 Myr, from 30 to 100 Myr in steps of 10 Myr and steps of 100 Myr after that). The redshift range varied from z=0 to 9 in redshift bins of $\Delta z=0.1$. The maximum age possible in each redshift is limited to the age of the Universe at that redshift. 

We divide galaxy models into three different categories: The dusty starburst galaxies with an exponentially declining star formation history, moderate to high values of extinctions and relatively young ages (measured from the onset of star formation); the elliptical galaxies with no dust extinction and old ages; and the passively evolving post-starburst galaxies with moderate ages and low extinction values. The dusty starburst galaxies could also have an older age but with a new burst of star formation. The post-starburst galaxy population is considered to be the phase after the initial starburst in galaxies when the more massive stars leave the main sequence after few hundred million years and is considered to be the phase before galaxies evolve into the early types at the lower redshifts with very old ages. The parameters used for the model evolutionary track are listed in Table 1.

\begin{table}
\centering
\caption{Parameters used for the model evolutionary tracks.}
\begin{tabular}{*{5}{c}}
\hline
\hline
Models & Age & Extinction & $\tau$ & Redshift \\
 & (Gyr) & E(B-V) & (Gyr) &  \\
\hline
Post Starburst & 0.3-1.5 & 0.0-0.2 & 0.0 & 0.0-9.0 \\
Dusty Starburst & 0.005-0.1 & 0.4- 0.9 & 0.0-0.2 & 0.0-9.0 \\
Elliptical & 1.5-2.5 & 0.0 & 0.0 & 0.0-3.0 \\
\hline
\end{tabular}
\end{table}

We integrated the WFC3 near-IR and HAWK-I $K_s$ band filter response functions over the model SEDs and found the corresponding magnitudes in the desired filters after incorporating the stellar and intergalactic extinctions. Changes in the color evolutionary tracks with redshift were then calculated. The degeneracies in the stellar population model parameters such as age, redshift and extinction, can also lead to red ($H_{160}-K_s$) colors. It is important to use more than one color to break the degeneracy between the high redshift relatively quiescent galaxies (BBGs) and the dusty starburst systems. Therefore, in addition to the main ($H_{160}-K_s$) colors we used the WFC3 ($J_{125}-H_{160}$) and ($Y_{105}-J_{125}$) and also non-detection conditions in the U and $B_{435}$ bands in order to minimize the fraction of the contaminants. The region on the color-color plots where BBG candidates are located is shown as the grey shaded area in Figure 2. This is adopted to minimize the number of contaminants.

Therefore, for the post starburst galaxies on the ($Y_{105}-J_{125}$) vs ($H_{160}-K_s$) plot, we choose the following selection criteria:

\begin{subequations}\label{grp}
\begin{align}
    &S/N(U) < 2 ,\label{first}\\
    &S/N(B_{435}) < 2 ,\label{second}\\
    &(Y_{105}-J_{125}) < -3.44+3.40 \times (H_{160}-K_s),\label{third}\\
    &(Y_{105}-J_{125}) > 1.03-0.67 \times (H_{160}-K_s),\label{fourth}\\
    &(Y_{105}-J_{125}) > -0.83+0.67 \times (H_{160}-K_s).\label{fifth}
\end{align}
\end{subequations}
\smallskip

and for the ($J_{125}-H_{160}$) vs ($H_{160}-K_s$) plot, we choose: 

\begin{subequations}\label{grp}
\begin{align}
    &S/N(U) < 2 ,\label{first}\\
    &S/N(B_{435}) < 2 ,\label{second}\\
    &(J_{125}-H_{160}) < -2.48+2.80 \times (H_{160}-K_s) ,\label{third}\\
    &(J_{125}-H_{160}) > 1.70-(H_{160}-K_s) ,\label{fourth}\\
    &(J_{125}-H_{160}) > -0.16+0.43 \times (H_{160}-K_s) .\label{fifth}
\end{align}
\end{subequations}

\begin{figure*}
\centering
\leavevmode
\includegraphics[scale=0.42]{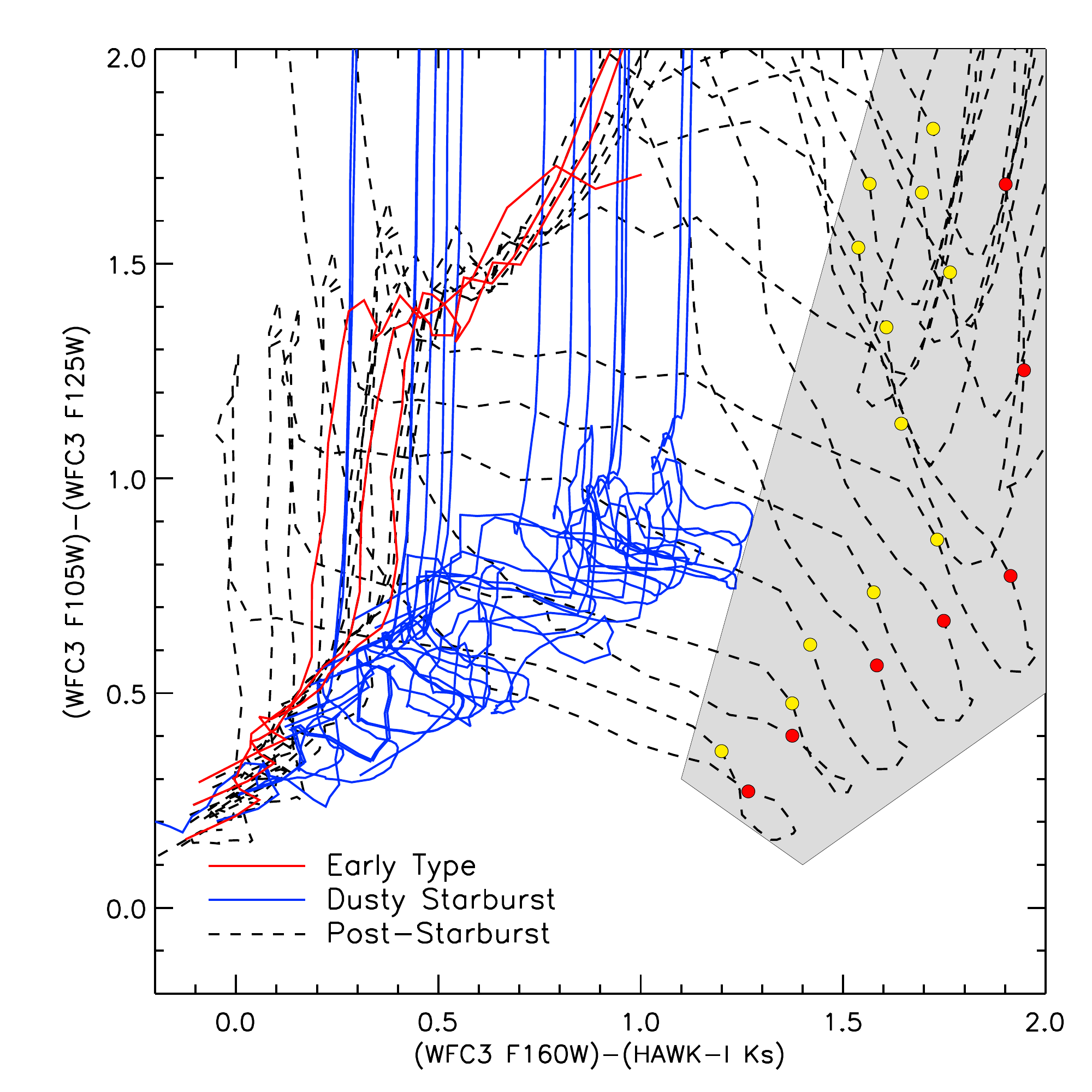}
\includegraphics[scale=0.42]{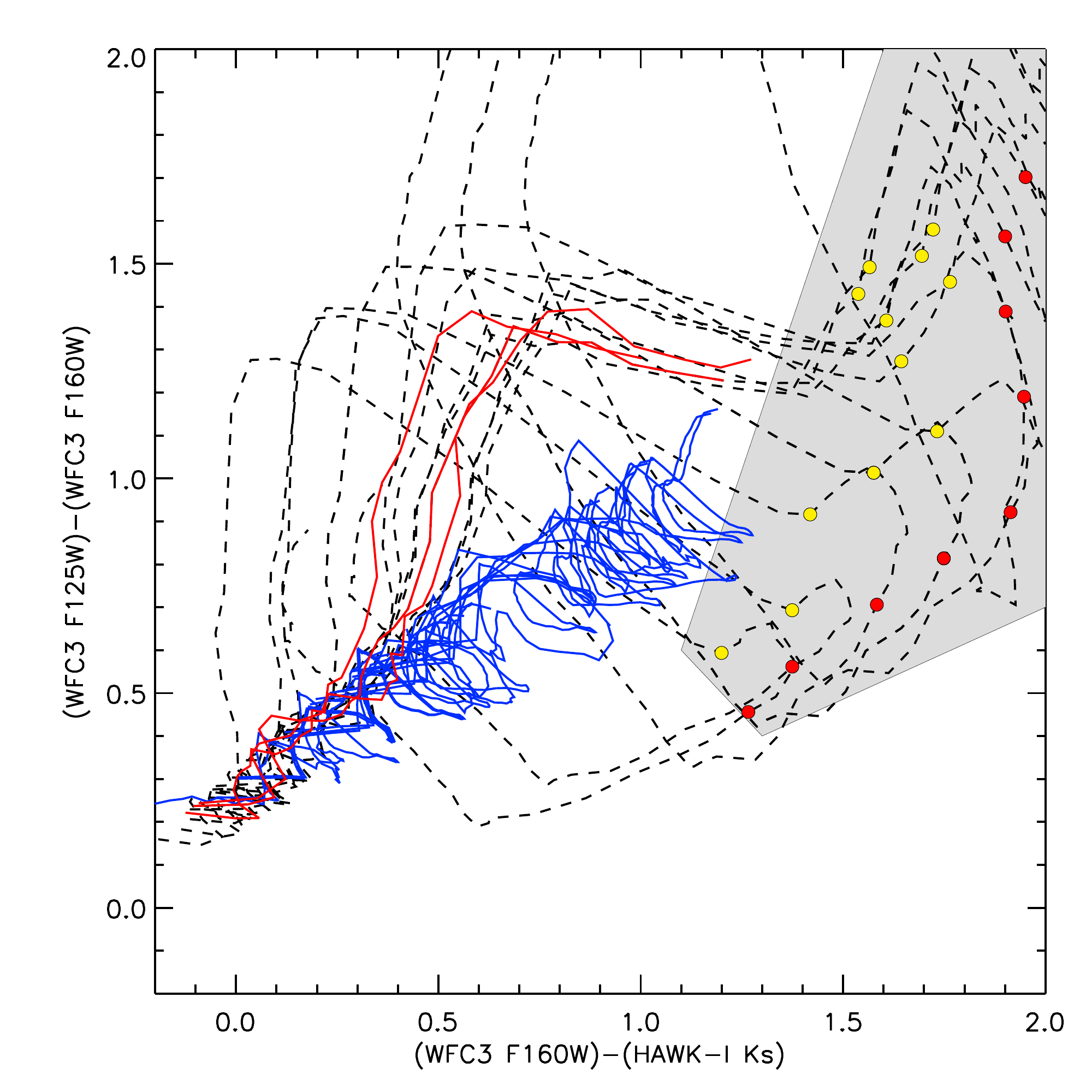}
\caption{$(Y-J)$ versus $(H-K_s)$ (left) and $(J-H)$ versus $(H-K_s)$ (right) evolutionary tracks from BC03 models. In both plots solid red, solid blue and dashed black evolutionary tracks represent elliptical, dusty starburst and post starburst galaxies respectively. The parameters used to generate the model tracks are listed in Table 1. The yellow and red circles on the post-starburst galaxy tracks correspond to redshifts 3.0 and 4.5 respectively. The BBG selection criteria is identified by the grey shaded area on the plots.}
\end{figure*}

Equations 1a-2e correspond to the shaded region in Figure 2 and are taken as our BBG selection criteria. In addition to the color location within the diagram, we imposed an additional constraint. The selected BBGs should also not be detected in the U and ACS $B_{435}$ bands, as in the redshift range of our interest the Lyman break falls long-ward of these filters. These additional conditions help in removing the lower redshift contaminants from the sample.

To further test the validity of the selection criteria, we generated post-starburst and dusty starburst evolutionary tracks as a function of the age of the galaxy. Analysis of these age-dependent evolutionary tracks recovers the same selection criteria that we originally identified for post-starburst galaxies. Also we find that the main source of contamination is from very dusty starburst galaxies with strong nebular emission lines (and not lower redshift early type galaxies).

\subsection{Simulations}

In order to assess the robustness of our color selection criteria we generated a WFC3 $H_{160}$-band selected mock catalog. The simulated catalog resembles the GOODS-S field as closely as possible. It has the same filters with the same wavelength, coverage area and photometric depth. We use the observed luminosity functions from \citealp {Dahlen2007} and randomly extracted galaxies with different absolute magnitudes from the luminosity function and randomly assign redshifts to each extracted galaxy. We then assigned spectral types to the simulated galaxies in the three main categories of early types, late types and starbursts. We then randomly assign photometric errors to the simulated galaxies and we add extinction spanning the range $E_{B-V}=0-0.9$, based on the galaxies assigned spectral types. 

The simulated catalog contains sources to a detection limit of 3 $\sigma$ in F160W. We used the observed Coleman, Wu and Weedman (CWW) spectral templates \citep {Coleman1980} to calculate the K-correction. Using the assigned redshift to calculate the distance modulus, we measured the apparent magnitude in each of the filters for the simulated galaxies. The result is a mock galaxy catalog with apparent magnitudes in the same filters as in the GOODS-S, redshifts, spectral types and extinction values. 

Figure 3 shows the simulated color-color plots overlaid on the BBG selection criteria developed in Section 3.1. The green points represent the post-starburst galaxy population in the mock catalog that are in the redshift range z $\simeq$ 3-4.5. We also show the simulated data for the early type and dusty starburst galaxies as red and blue points respectively. The simulations show that there is very little contamination from the elliptical galaxies while the dusty starburst galaxies still contaminate the BBG sample. We will address this issue later in Section 4 when we fit the spectral energy distribution of the BBG selected candidates to galaxy models. 

\begin{figure*}
\centering
\leavevmode
\includegraphics[scale=0.42]{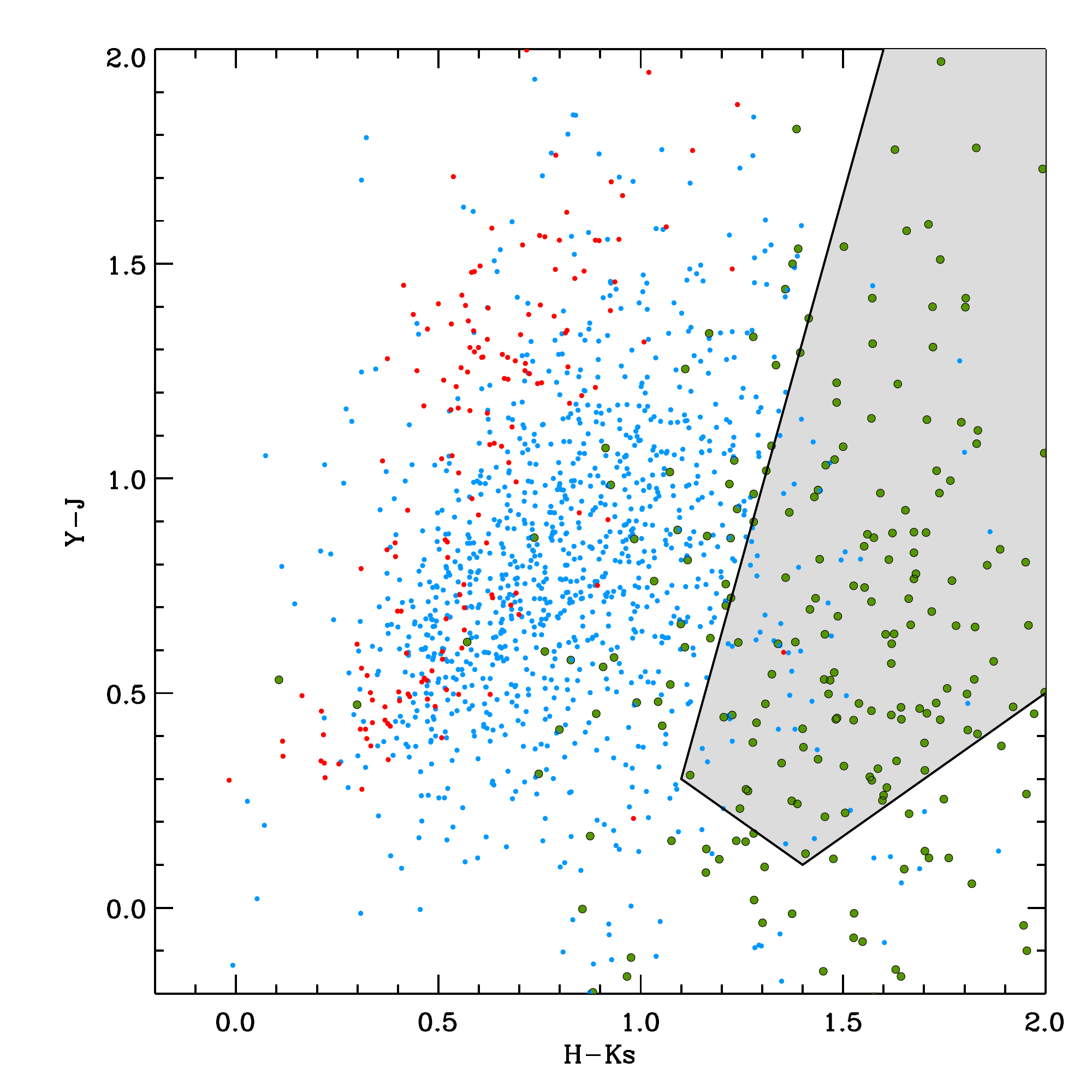}
\includegraphics[scale=0.42]{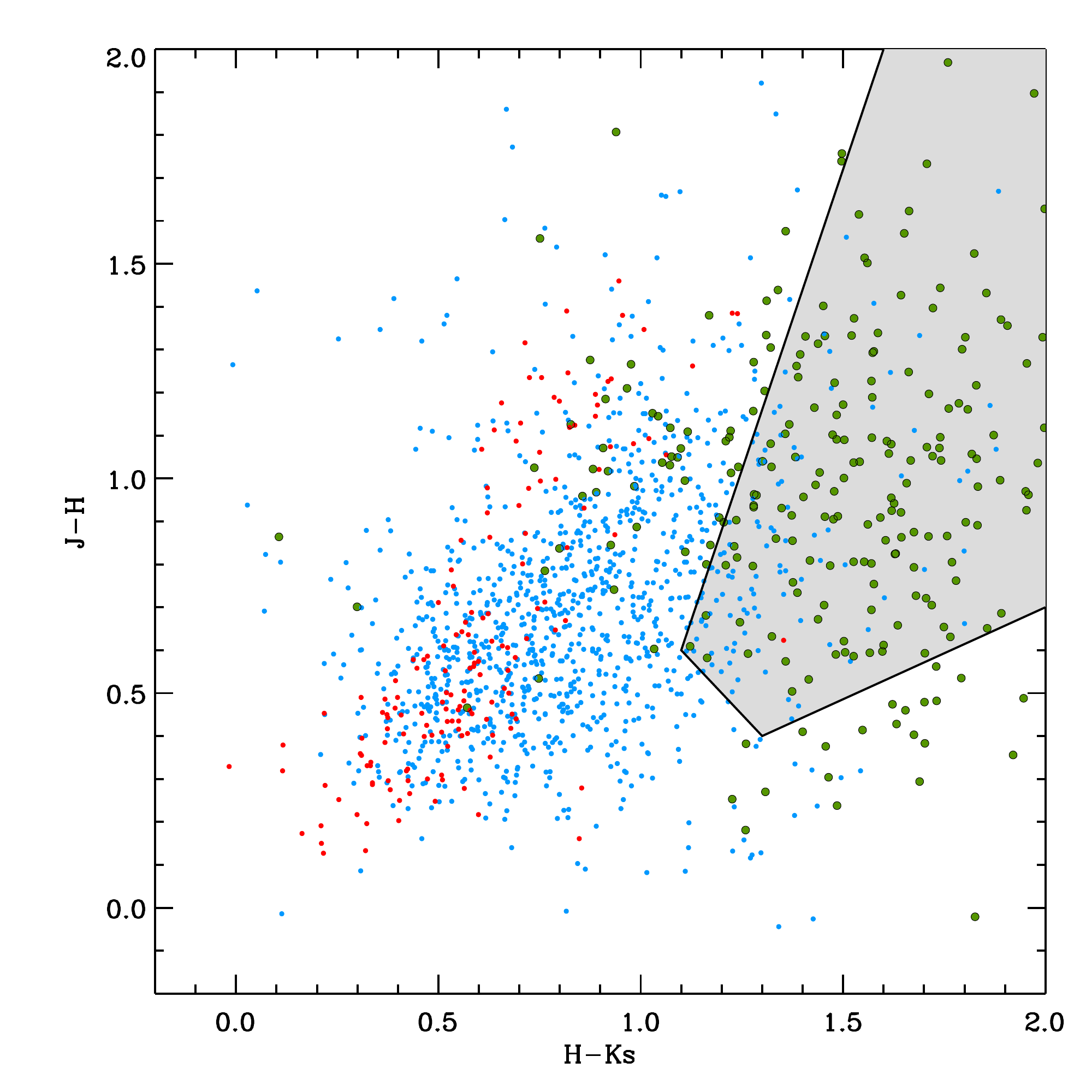}
\caption{Simulation color-color diagrams. The blue, red and green data points correspond to elliptical, dusty starburst and post starburst galaxies from our simulation respectively with the BBG selection shown as grey shaded area. Although we get very little contamination from the lower redshift early type galaxies (red points), the dusty star-forming systems could still contaminate the sample (blue points).}
\end{figure*}

\subsection{Candidate Identification from the color-color Criteria}

We applied the BBG selection criteria derived in Section 3.1 to the deep WFC3 $H$-band selected TFIT catalog, setting a 10$\sigma$ detection requirement in the WFC3 $H$-band for the catalog. We apply the ($Y_{105}-J_{125}$) vs ($H_{160}-K_s$) and ($J_{125}-H_{160}$) vs ($H_{160}-K_s$) criteria on the WFC3 $H$-band selected catalog. We find a total of 24 sources in the combined two selection criteria identified as BBG candidates. Figure 4 shows the color selection criteria as the shaded region along with data from the $H$-band selected TFIT catalog. Sources that are within the shaded region but did not satisfy the U and/or B bands non-detection conditions were not included. The SEDs of all the identified candidates show pronounced Balmer breaks between the WFC3 H and HAWK-I $K_s$ bands, as expected from the color selection. 

\begin{figure*}
\centering
\leavevmode
\includegraphics[scale=0.42]{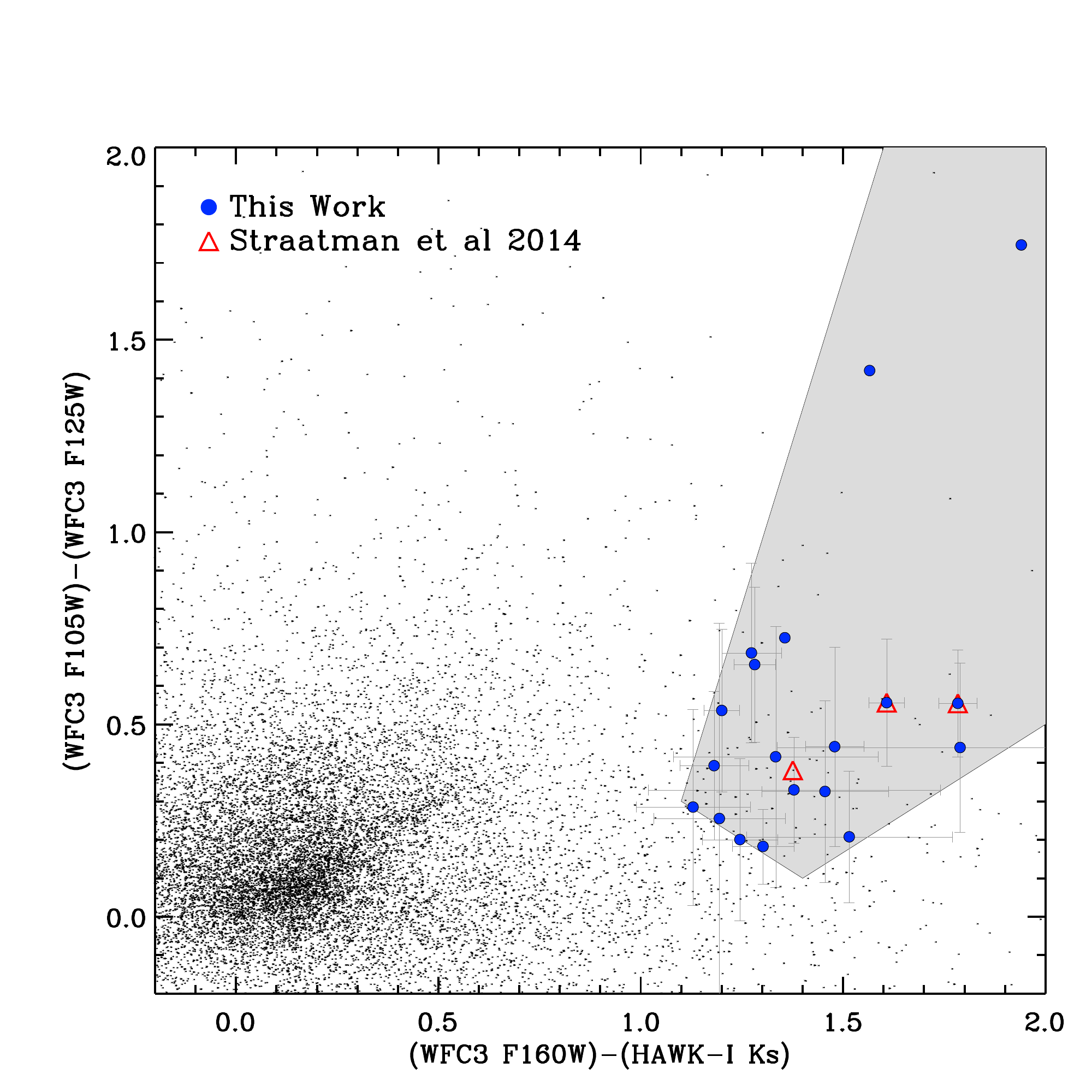} 
\includegraphics[scale=0.42]{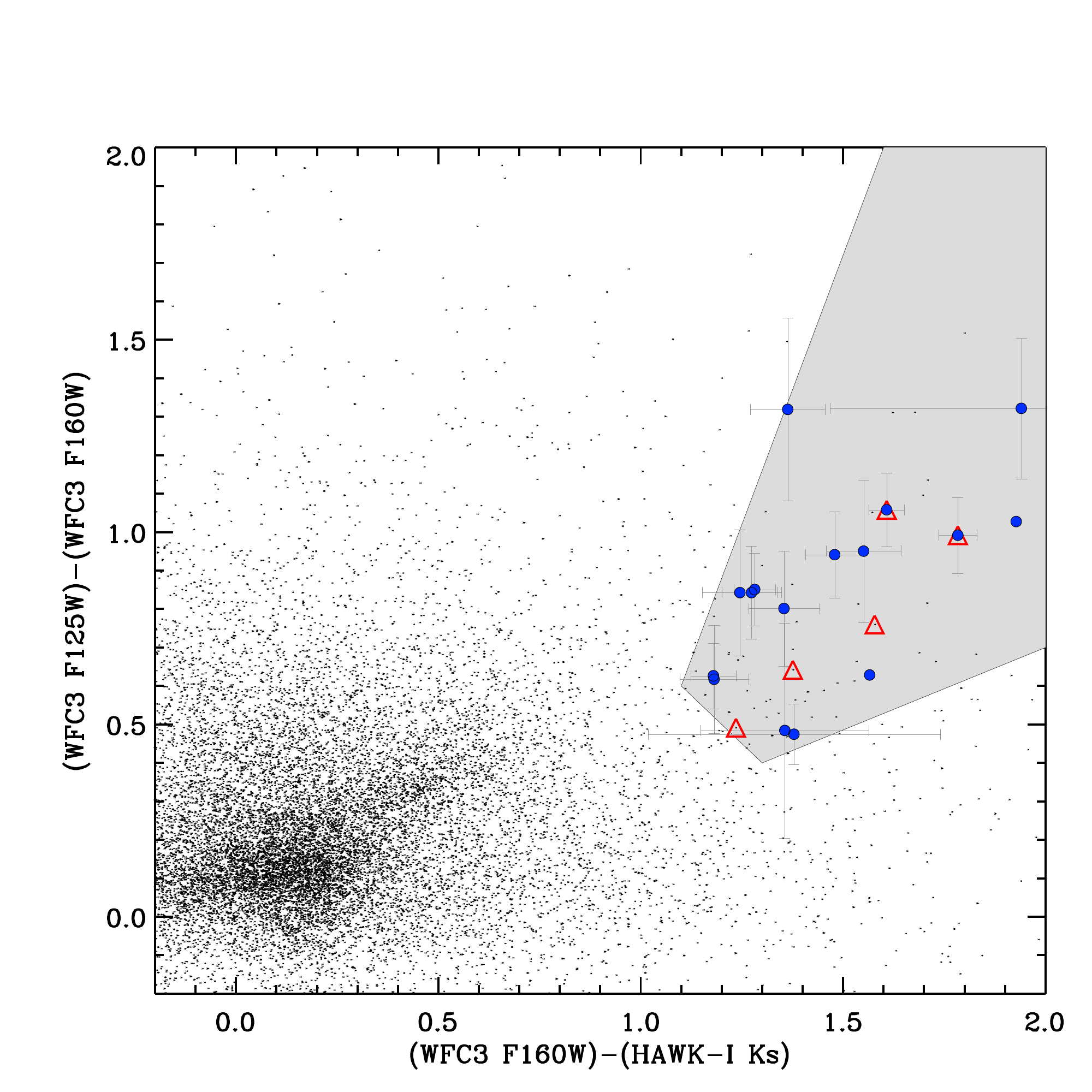}
\caption{Left: The $(Y-J)$ vs $(H-K_s)$ selection criteria (Eq. 1) applied to the CANDELS WFC3 $H$-band selected TFIT data (black data points). Right: Plot of $(J-H)$ vs $(H-K_s)$ (Eq. 2). Blue circles correspond to the selected candidates that lie within the selection criteria and are not detected in the U and B bands (i.e., are at $z \sim 3$ and $z \sim 4$). The red triangles are the UVJ selected passive systems by \citealp {Straatman2014}. The final candidates are from the two combined color-color plots.}
\end{figure*}

One widely used selection criteria for distinguishing quiescent galaxies from star forming systems is by using the rest-frame U, V and J in a (U-V) vs. (V-J) color plot (the so called UVJ selection; \citealp {Labbe2005, Wuyts2007, Williams2009}). The bimodality between dusty star forming and quiescent galaxies in the UVJ selection becomes less prominent at $z>2$ mostly because of photometric redshift uncertainties and weaker fluxes \citep {Williams2009}. Figure 5 shows the distribution of the near infrared BBG selected sources on a UVJ color-color plot. We see from the Figure that many of the candidates are at the boundary of dusty star-forming and quiescent criteria or will be identified as dusty galaxies by this selection.

Recently \citealp {Straatman2014, Spitler2014} extended the UVJ selection of passive galaxies to $z \sim 3-4$ using medium band data. The UVJ selected quiescent candidates by \citealp{Straatman2014} are plotted along with the BBG selected candidates in Figure 4. All the UVJ selected quiescent candidates fall on the BBG color selection criteria while not all of them are selected as BBG candidates because of the additional criteria imposed on the U and $B_{435}$ bands. 

\begin{figure}
\includegraphics[scale=0.5]{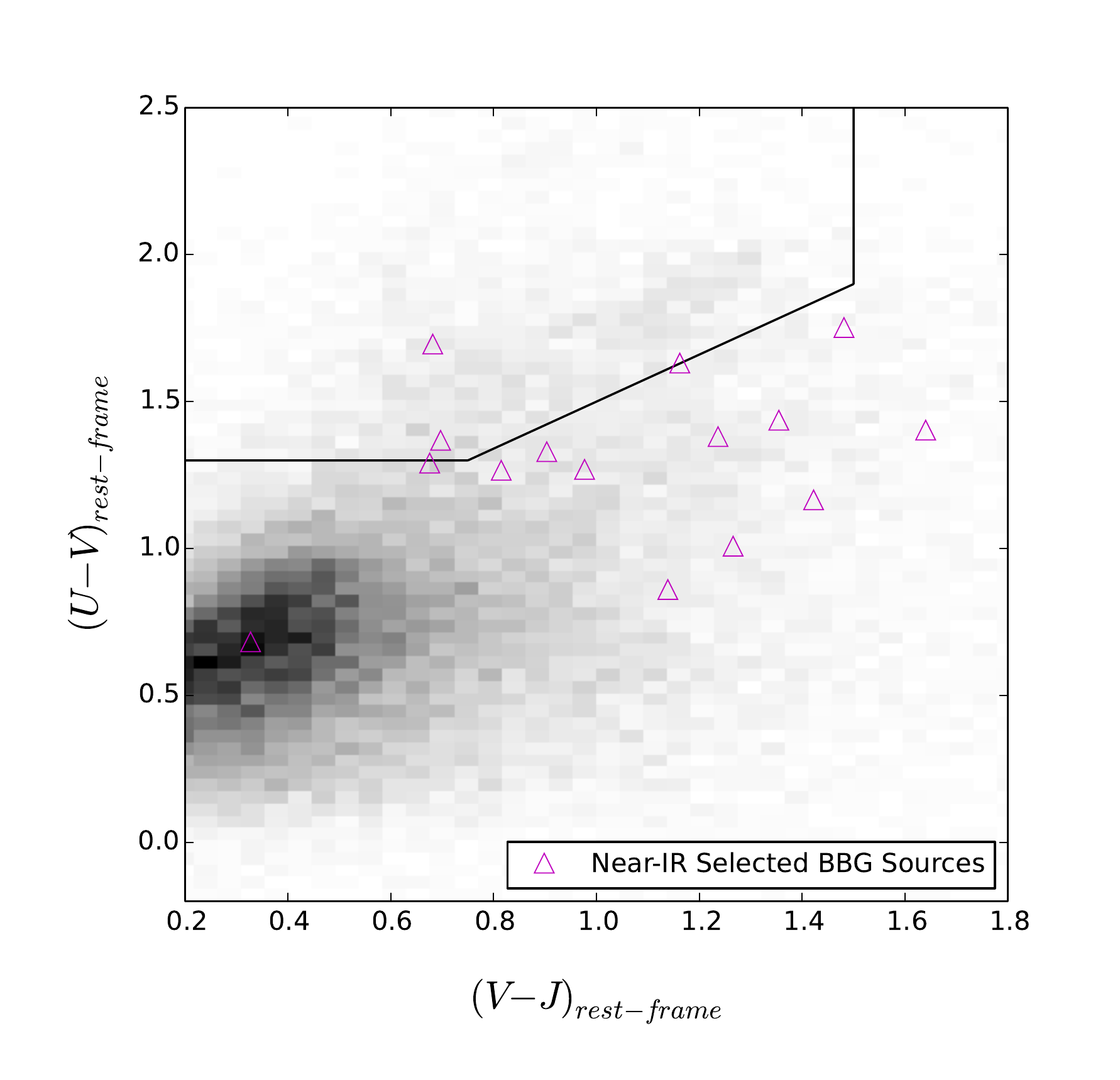}
\caption{Distribution of the near infrared color selected BBG sources on the rest-frame UVJ color selection criteria \citep {Williams2009}. The area on the plot shows the UVJ predicted location for quiescent galaxies. The BBG selected sources fall close to the boundary of dusty star forming and quiescent candidates.}
\end{figure}

\subsection{Effect of Nebular Emission on the BBG Selection}

The nebular emission lines can play a significant role in both the selection and SED fitting of the BBG candidates. The sensitivity of the final estimated parameters on the nebular emission will be discussed in Section 4 while, here we examine the effect of nebular emission on colors and hence on selection of the BBGs.

We account for the effect of nebular emission (both continuum and emission lines) following \citealp {Schaerer2009} and \citealp {Schaerer2010}. The strength of nebular emission depends mainly on the number of Lyman continuum photons, which is computed from stellar population synthesis models. Relative line intensities are taken from \citealp {Anders2003} and \citealp {Storey1995}, for a typical ISM ($n_e = 100\: cm^{−3}$, $T = 104^{\circ}$K). We assume that the Lyman escape fraction is equal to zero, which means that our models produce the maximum theoretical strength for nebular emission (under our assumptions about the ISM). Additionally, we also apply the same attenuation between nebular and stellar emission.

Figure 6 shows an example of the observed photometry for one of our BBG candidates at z= 3.6, along with fitted template SEDs without nebular emission (green), showing the Balmer break at $\sim 2 \mu$m, and with nebular emission (blue). This reveals how redshifted nebular emission could disguise itself as a Balmer break feature. The infrared broadband photometry of the candidates could also be affected by the nebular emission lines. In particular, in the redshift range covered in this study, the H$\alpha$ line at rest-frame 6563\AA\ could potentially contaminate the IRAC 3.6 $\mu$m flux at $z \sim 3.8$. In order to further study the effect of nebular emission lines on the SED fit for our candidate galaxies, we measured the IRAC ($m_{3.6}-m_{4.5}$) color of BBG sample as the presence of the H$\alpha$ line is only expected to affect the 3.6 $\mu$m band as it shifts to this band at $a \sim 3.8$. For our BBG sample, there are four galaxies that have both IRAC 3.6 $\mu$m and 4.5 $\mu$m detections at the redshift range affected by this emission. Figure 7 shows the IRAC colors of the four BBG candidates for which H$\alpha$ emission lines shift to the IRAC 3.6$\mu$m. For three of these candidates, it is clear that the observed infrared colors are better represented by model SEDs that include nebular emission in the fit, accounting for the H$\alpha$ contribution. Post-starburst galaxies are relatively old systems and we do not expect to see strong H$\alpha$ emission in their spectra. The candidates that are affected by this feature are relatively younger. Figure 7 implies that the absence of correction for nebular emission in the redshift range of interest here would lead to redder ($m_{3.6}-m_{4.5}$) colors, which would mimic the Balmer break and hence, overestimate their stellar mass. 

In order to better quantify the effect of nebular emissions, we also generated color evolutionary tracks for the post-starburst galaxies including contribution from nebular emission and recovered the same color selection criteria as for the model tracks in Figure 2, where no nebular emissions were incorporated. The effect of nebular emission on evolutionary tracks is most noticeable for dusty star-forming galaxies. For the highly dusty star-forming systems, inclusion of nebular emission would further shift them into the BBG selection domain, further contaminating the sample. Most of the contamination in the sample comes from this heavily dust obscured, young and star-forming population.

\begin{figure}
\includegraphics[scale=0.45]{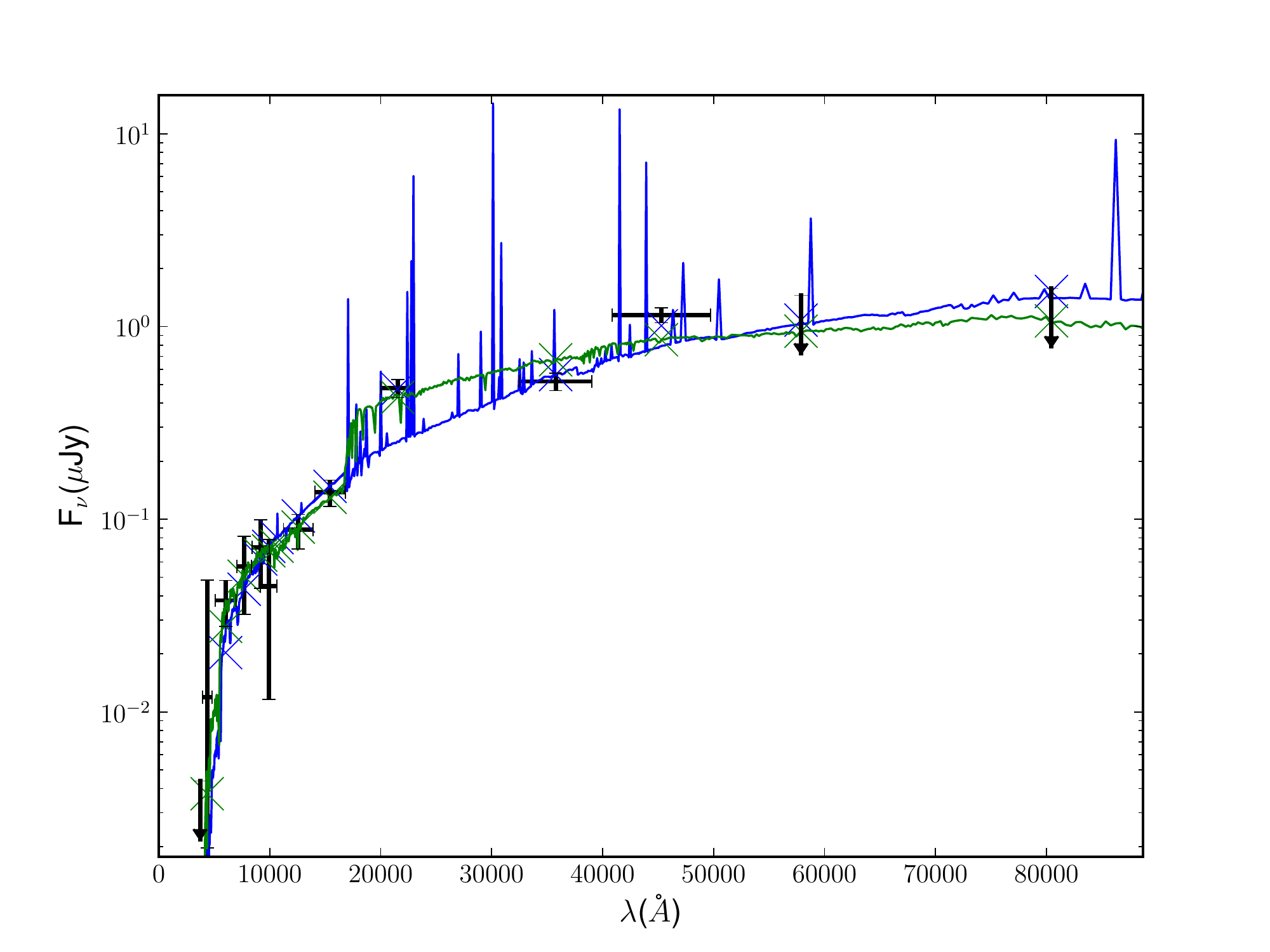}
\caption{Example of an SED fit to a BBG candidate at z=3.60 without nebular emission (green) and with nebular emission (blue). This shows the extent to which nebular emission could mimic the Balmer break feature shifted to near IR.}
\end{figure}

\vspace{10 mm}

\begin{figure*}
\centering
\leavevmode
\includegraphics[scale=0.42]{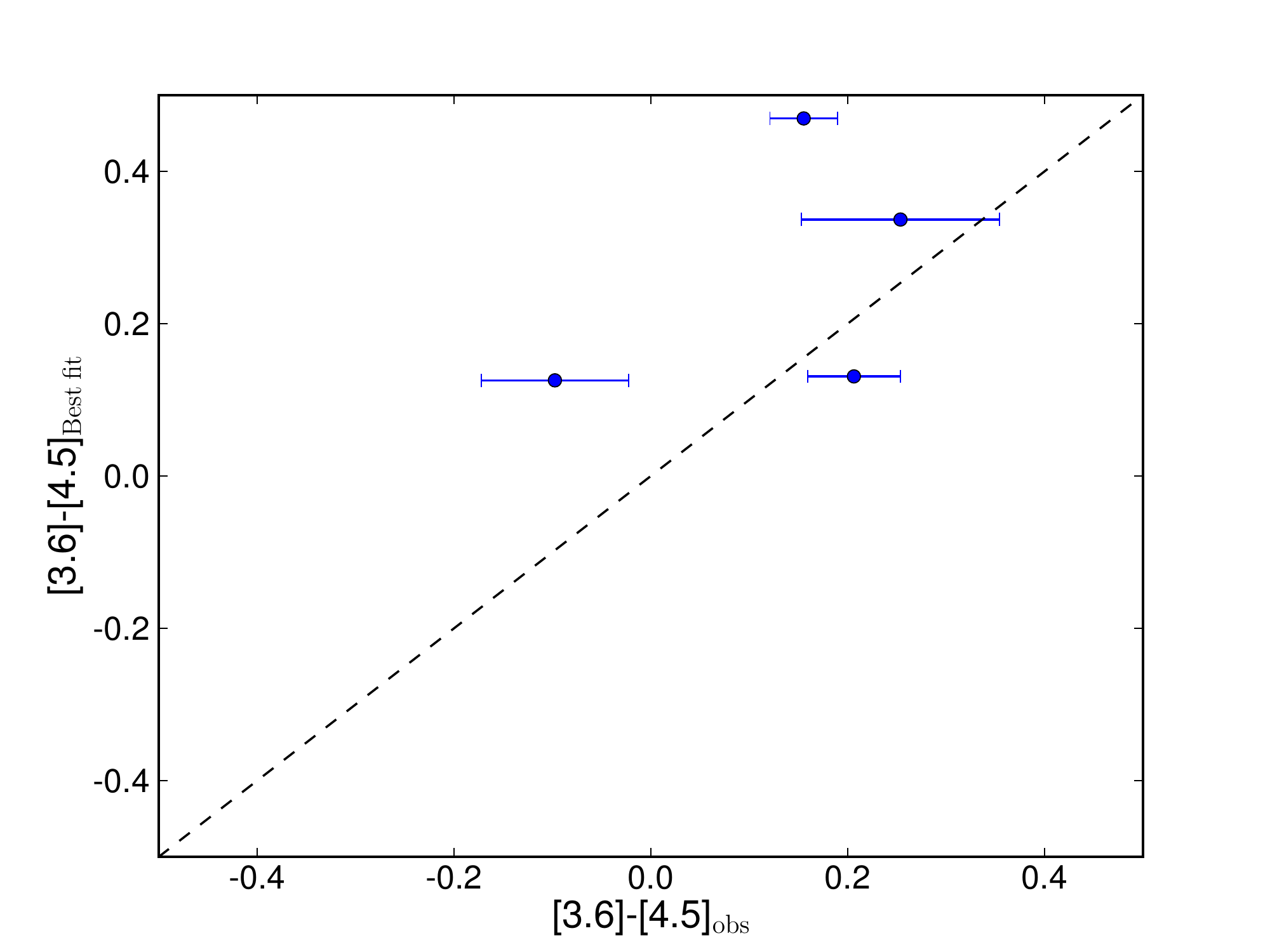} 
\includegraphics[scale=0.42]{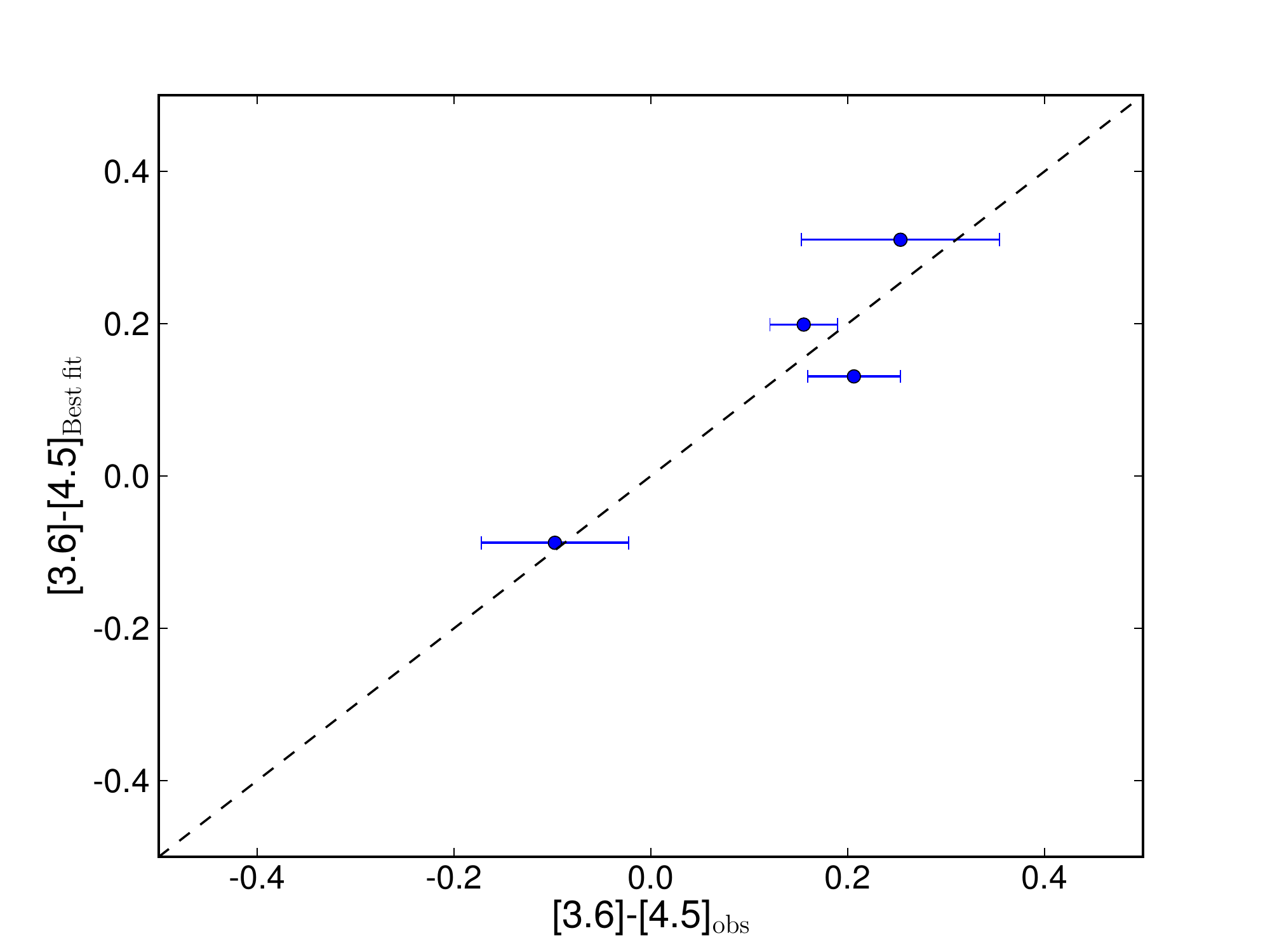}
\caption{Left: The observed IRAC colors of the 4 BBG candidates with strong contribution from the H$\alpha$ shifted to the infrared pass-bands are plotted against the best fit model color with no nebular emission contributions incorporated in the fit. Right: Same as the left plot but with nebular emission added to the fit.}
\end{figure*}

\section{Measurements of the Observable Parameters}

In this Section we study the nature of the BBG candidates by measuring their physical properties (i.e. stellar mass, age etc.). We use this to examine if the candidates fit the expected requirements for the BBGs. The photometric redshifts and physical parameters of the BBG candidates, selected from the criteria in Section 3.1, are inferred from fitting the observed SEDs with a BC03 generated template library using a modified version of the SED fitting code HyperZ \citep {Bolzonella2000}. In the SED fitting code, the redshift is a free parameter that is allowed to vary from z=0 to z=9 for fits with and without nebular emission. We consider solar metallicity and allow the extinction to vary from $0<A_v<3$. In the following we will discuss the SED fitting results of BBG selected sample.

\subsection{Effect of Nebular Emission on Parameter Estimation}

Adding nebular emission to the SED fitting is expected to lower the age and stellar mass estimate for the candidates that are affected by these emissions as it would reduce the mass to light ratio of these systems.

Figure 8 compares the stellar mass estimates for the BBG sample with and without nebular emission. Here, the blue filled circles correspond to the BBG candidates for which the SED fit, after adding nebular emission, gives a lower age estimate (by more than 0.5 dex). There is excellent agreement between the stellar mass estimates for the BBG candidates for which the difference in age estimate is small ($\Delta log(age)<0.5$ ). Figure 9 shows the age distribution of the color-selected BBG candidates that are affected by the nebular emission (blue filled circles in Figure 8). Here, including nebular emission in the fit significantly affects their parameter estimation as seen from the distributions. Out of the 24 BBG candidates selected from color-color plots, 8 show strong nebular emission effect in their SED fit and were removed from the BBG sample as this causes false identification. The three candidates mentioned in Section 3.4 as having strong H$\alpha$ contamination are among these 8 galaxies. The strong nebular emission line galaxies are identified based on having a better $\chi^2$ fit with nebular emission. Figure 10 shows the postage stamp images of the final 16 BBG candidates in the multi-waveband. We see from this Figure that many of the BBG selected sources are relatively compact. This is in agreement with recent studies indicating that the first galaxies to become passive in the Universe are the massive compact systems \citep {Cassata2013, Williams2014}.

\begin{figure}
\includegraphics[scale=0.42]{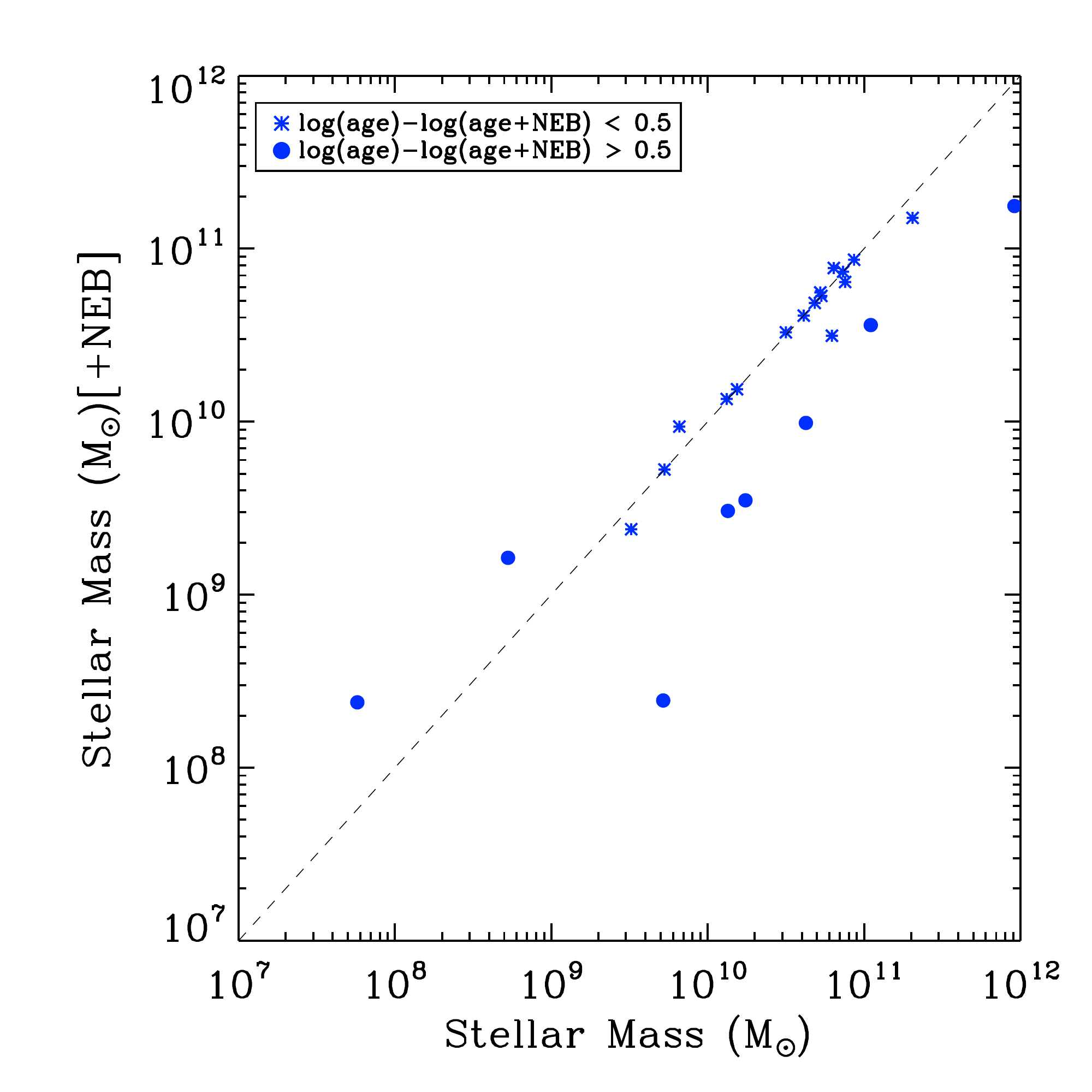}
\caption{Comparison of the stellar mass values for BBGs estimated with and without including nebular emission contribution. The blue filled circles correspond to sources with an age estimate difference of larger than 0.5 dex in fits with and without nebular emission.}
\end{figure}

\begin{figure}
\includegraphics[scale=0.35]{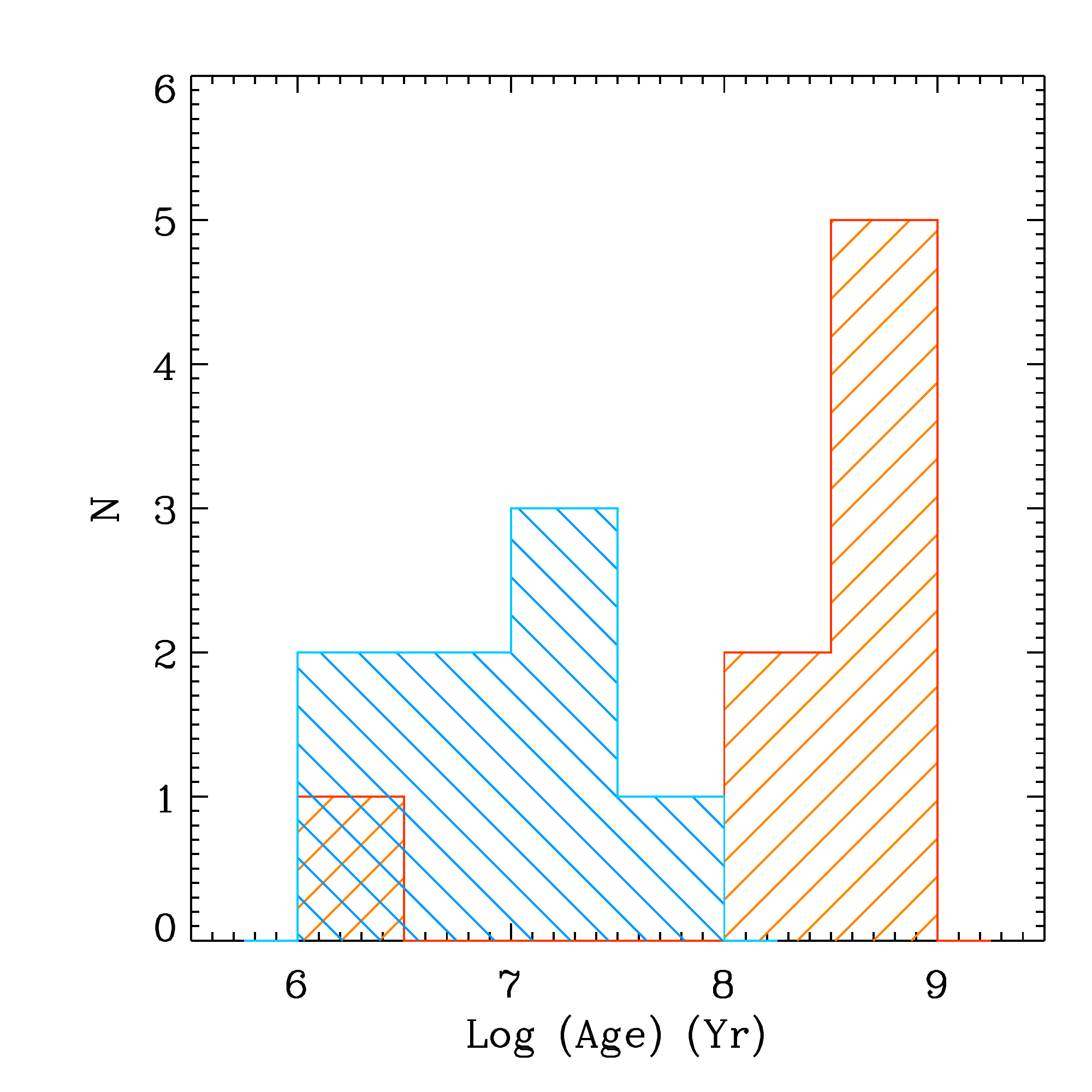}
\caption{The age distribution of BBGs shown with filled symbols in Figure 8. The red histogram shows the age distribution of these sources without nebular emission added to the SED fit and the blue histogram is the age distribution with the nebular emission effect added to the fit. These are the galaxies for which both their age and stellar masses are sensitive to the inclusion of nebular emission lines. Clearly including nebular emission lines leads to younger ages and lower stellar mass estimates.}
\end{figure}

\begin{figure*}
\includegraphics[scale=0.80]{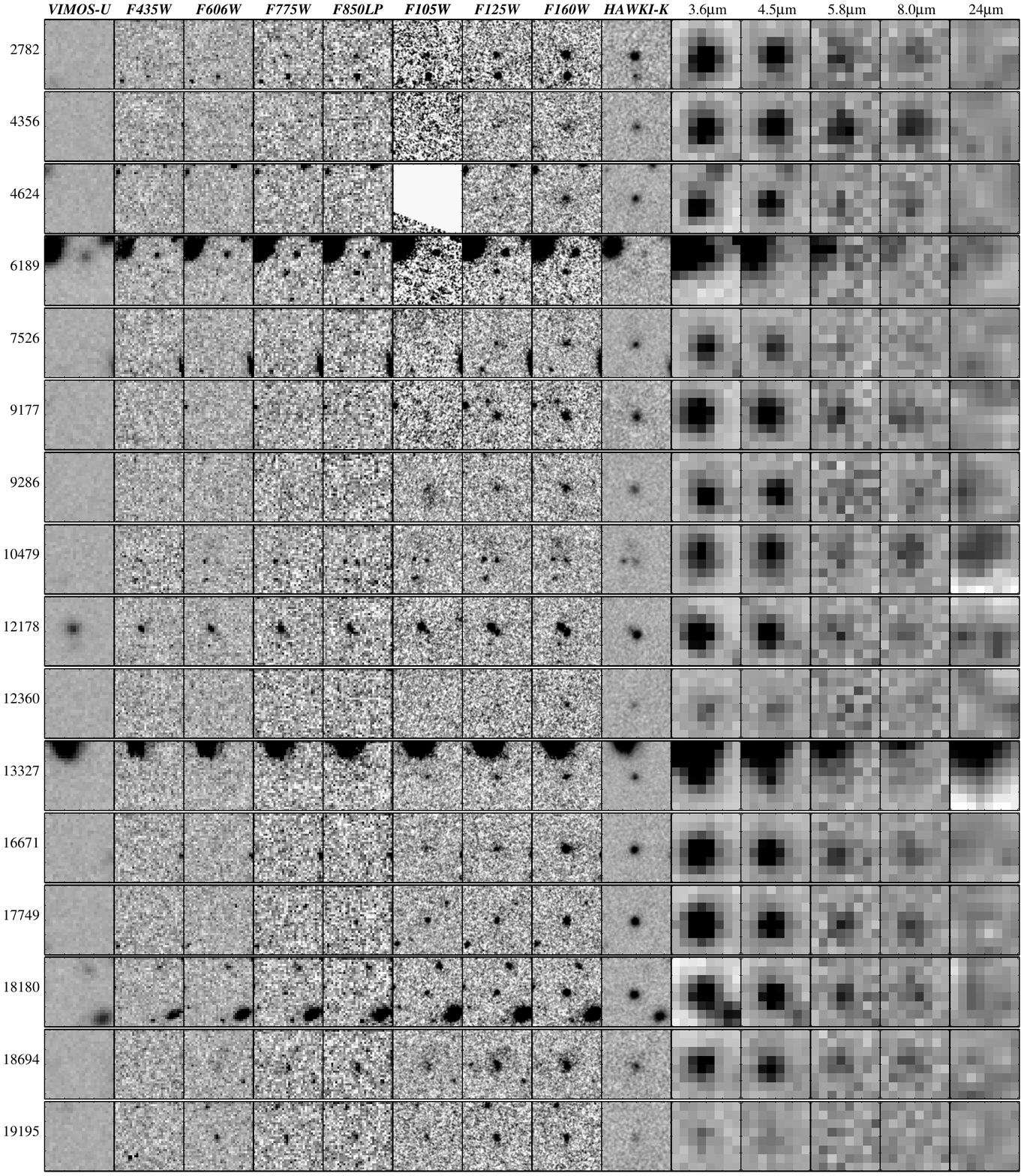}
\centering
\leavevmode
\caption {Postage stamp images of the BBG candidates. From left to right, the VLT/VIMOS U band, ACS optical, WFC3 near-IR, VLT/HAWKI $K_s$, Spitzer IRAC and MIPS 24 micron bands. The candidates are very faint in the bluer bands and get brighter in the redder bands. Three of the candidates are marginally detected in the MIPS 24 $\mu$m. The cutouts are $5^{\prime\prime} \times 5^{\prime\prime}$ in size.}
\end{figure*}

\subsection{Photometric Redshift} 

By fitting the SED for individual BBGs, we estimated the photometric redshifts for each of our candidates selected from the color criteria. The relatively quiescent galaxy candidates at $z \sim 3-4$ currently lack spectroscopic confirmation due to the absence of prominent emission features because of their passive nature which would not allow us to further check the robustness of the estimated photometric redshifts. Figure 11 shows the distribution of the photometric redshifts for the color selected BBG candidates. As expected, the BBG population has a redshift distribution that peaks between $3.0<z<4.5$. 

\subsection{Mass Measurement}

The stellar masses for the BBG candidates were measured by fitting their SEDs to templates generated from the population synthesis models. In Figure 11 we plot the distribution of their stellar masses. The distribution has a median at $log(M/M_{\odot}) \sim 10.6$, with the majority of the BBGs having large stellar masses. Mid-infrared observations by the Spitzer Space Telescope provide a direct measure of the stellar mass at the redshift range considered here \citep {Papovich2006, Yan2006}. Massive galaxies are brightest in the rest frame near-IR band and at the redshift range considered here, that roughly corresponds to the observed IRAC 4.5 $\mu$m. The accuracy of the selection criteria and the estimated parameters are therefore directly affected by the depth and quality of the IRAC data. Therefore, the very deep IRAC data at 3.6 $\mu$m and 4.5 $\mu$m are essential for the present study as they are probing red-ward of the Balmer break. As discussed in Section 2.4, the depth achieved in the SEDS data \citep {Ashby2013} gives a unique opportunity to constrain the SEDs of the most massive galaxies and to accurately measure the mass.

\subsection{Extinction}

We derive the extinction of the BBGs from SED fitting. The extinction is added to the templates used for the SED fitting in the range of $0<A_v<3$. Figure 11 shows the extinction distribution for the color selected BBGs. From our classification in Table 1 we see that the post starburst population has modest extinction values. Figure 11 shows that the majority of the BBGs have low to moderate extinction values. There are four color selected candidates that have relatively large derived extinctions of $E(B-V)>0.3$. These galaxies have relatively large SED inferred SFRs indicating that they could be part of the dusty galaxy population that contaminate the sample by replicating near IR colors of the BBGs. We will further discuss this population in Section 5.

\subsection{Age}

The ages of the BBG sample is plotted in Figure 11. Age is perhaps the most loosely constrained parameter in the SED fit. It is also the parameter that affected the near infrared colors of the post-starburst evolutionary tracks the most. The BBGs are evolved population due to their strong Balmer breaks (Table 1) and have relatively old ages ($\sim 10^{9} yrs$). Also as we discussed earlier, this is the parameter that is most significantly affected (along with the stellar mass) by the presence of nebular emission lines to the SEDs of the galaxies (Figures 8 and 9).  

\section{MIPS detected BBGs}

We now study the properties of our candidate BBGs as a function of MIPS detection at 24 $\mu$m. The detection in the Spitzer MIPS at the redshift of interest here can be interpreted as due to dust. At this redshift, the MIPS (at 24 $\mu$m) probes rest-frame 5-6 $\mu$m which can result from PAH feature i.e., star formation. We crossed-matched our BBG candidates selected from the color-color plots with the MIPS catalog for the GOODS-S \citep {Magnelli2011}, using a matching radius of 1.5 arcseconds. The matching radius is chosen to identify the MIPS sources while avoiding contamination from the neighbors. Out of 16 candidates identified as BBGs, 3 have a MIPS source associated with them. The bluest strong PAH band is at 6.2 $\mu$m and for the BBG candidates to be bright enough to be detected by MIPS 24 $\mu$m band (assuming that the photometric redshifts are correct) this would mean that the detection is likely to come from hot dust continuum, most likely from an AGN. One of the sources is detected in both X-ray and MIPS, which indicates the presence of an obscured AGN. We find that the three candidate galaxies that are detected in the MIPS 24 micron have moderate to high extinctions, with the candidate that is selected in both the MIPS and the X-ray having SED measured color excess ($E(B-V)$) of 0.6, indicating heavy dust obscuration. The candidates with MIPS and/or X-ray detections are listed in Table 2. 

\section{Comparison between the sources selected by the LBG and BBG techniques}

\begin{figure*}
\centering
\leavevmode
\includegraphics[trim=3.5cm 0cm 0cm 0cm,scale=0.25]{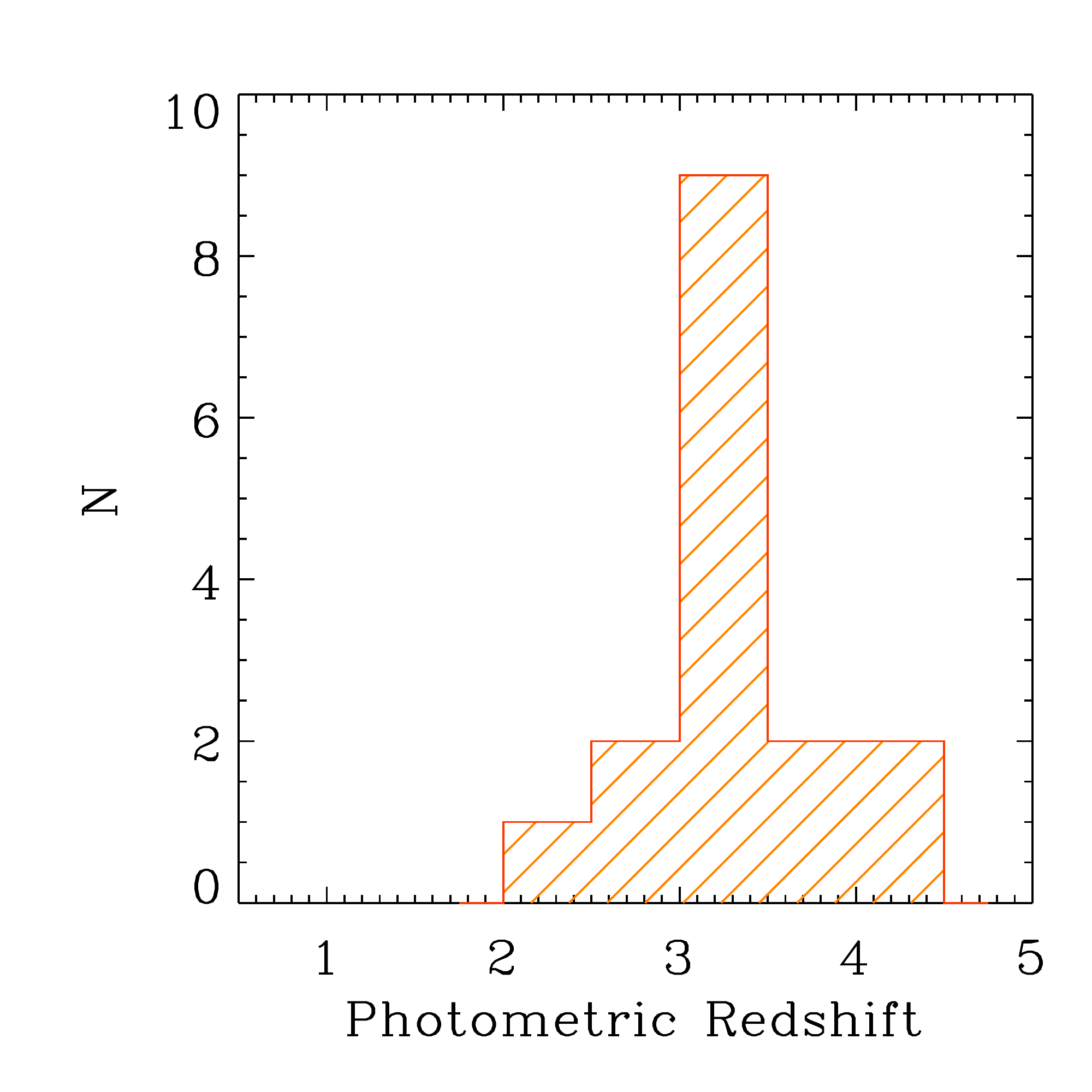}
\includegraphics[trim=3.5cm 0cm 0cm 0cm,scale=0.25]{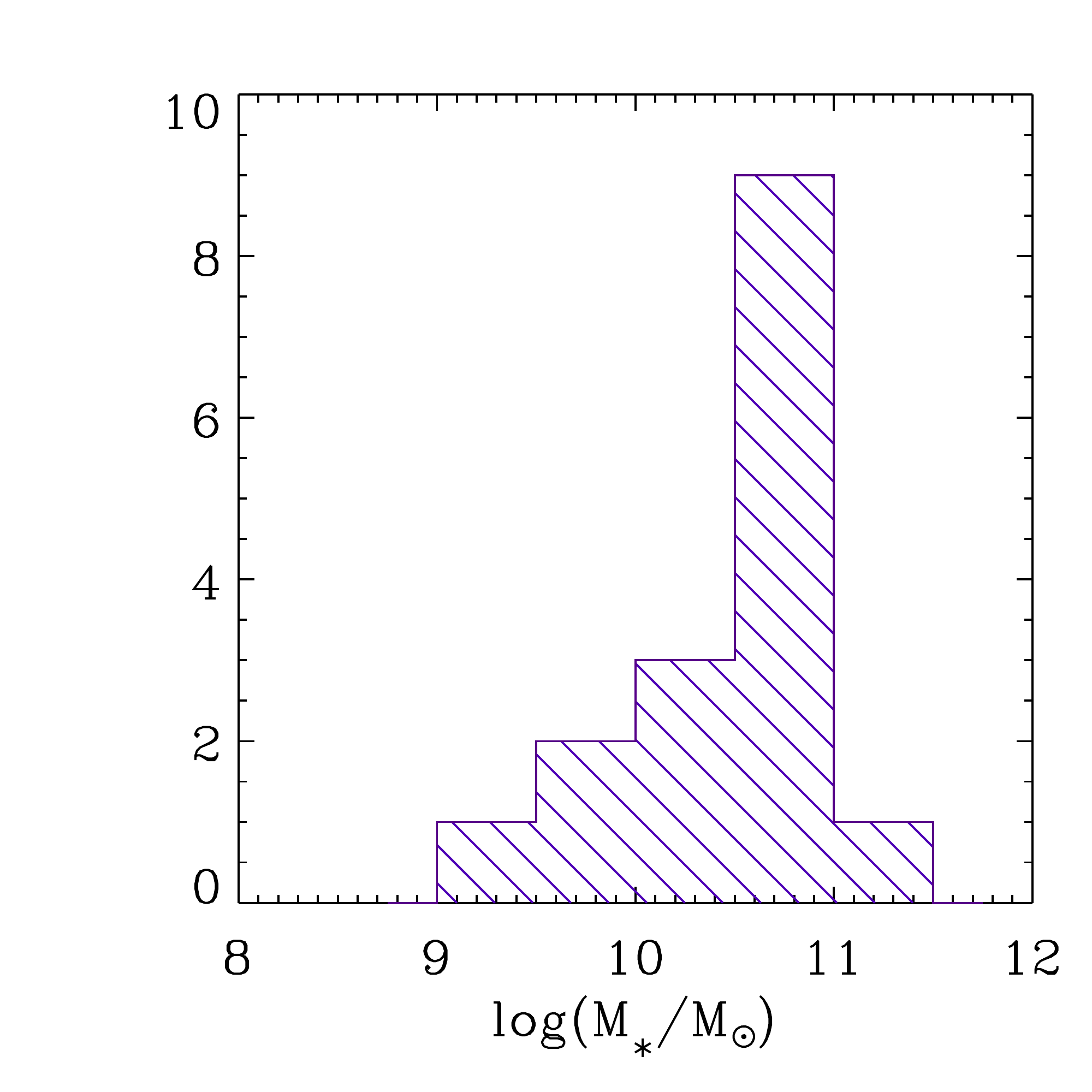} 
\includegraphics[trim=3.5cm 0cm 0cm 0cm,scale=0.25]{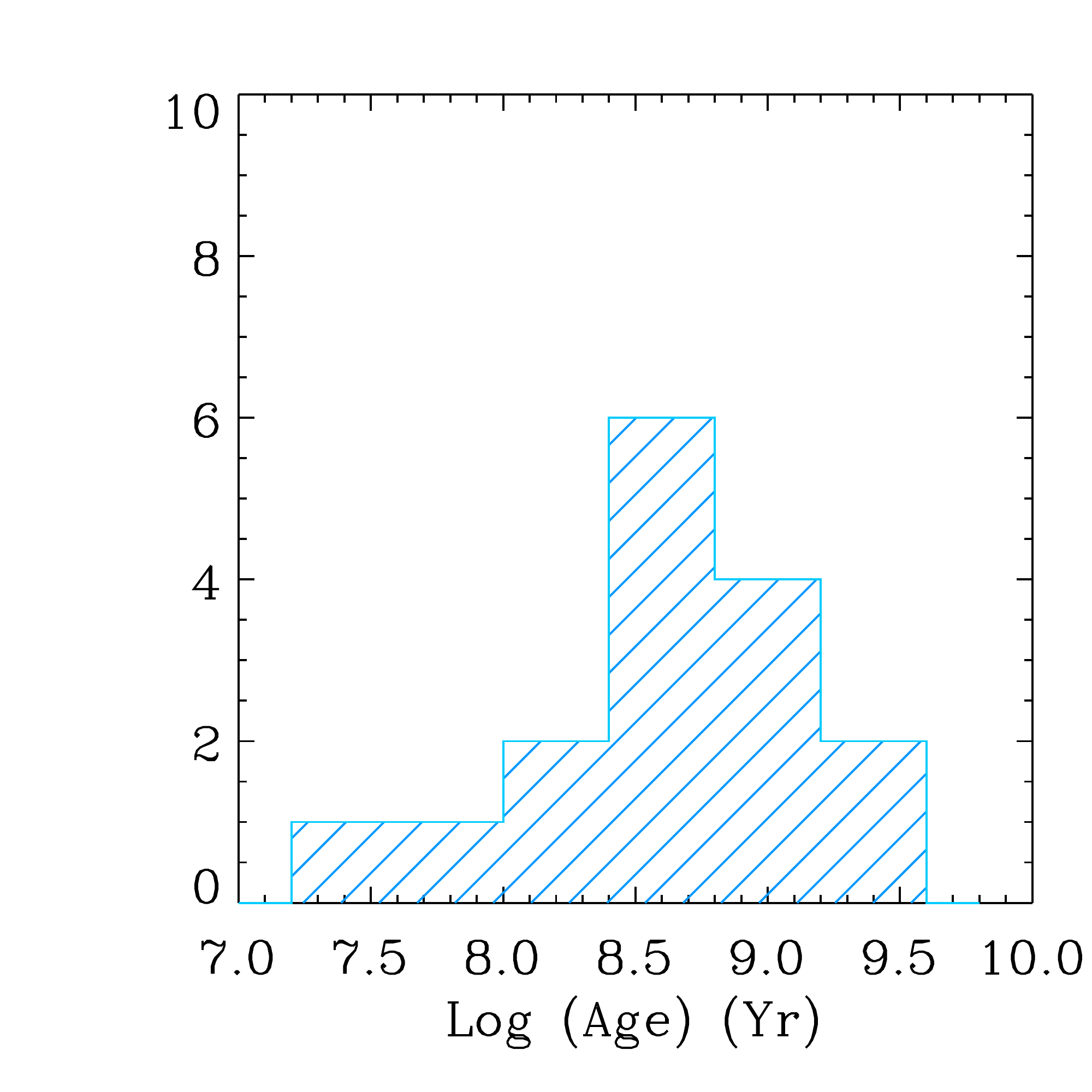}
\includegraphics[trim=3.5cm 0cm 0cm 0cm,scale=0.25]{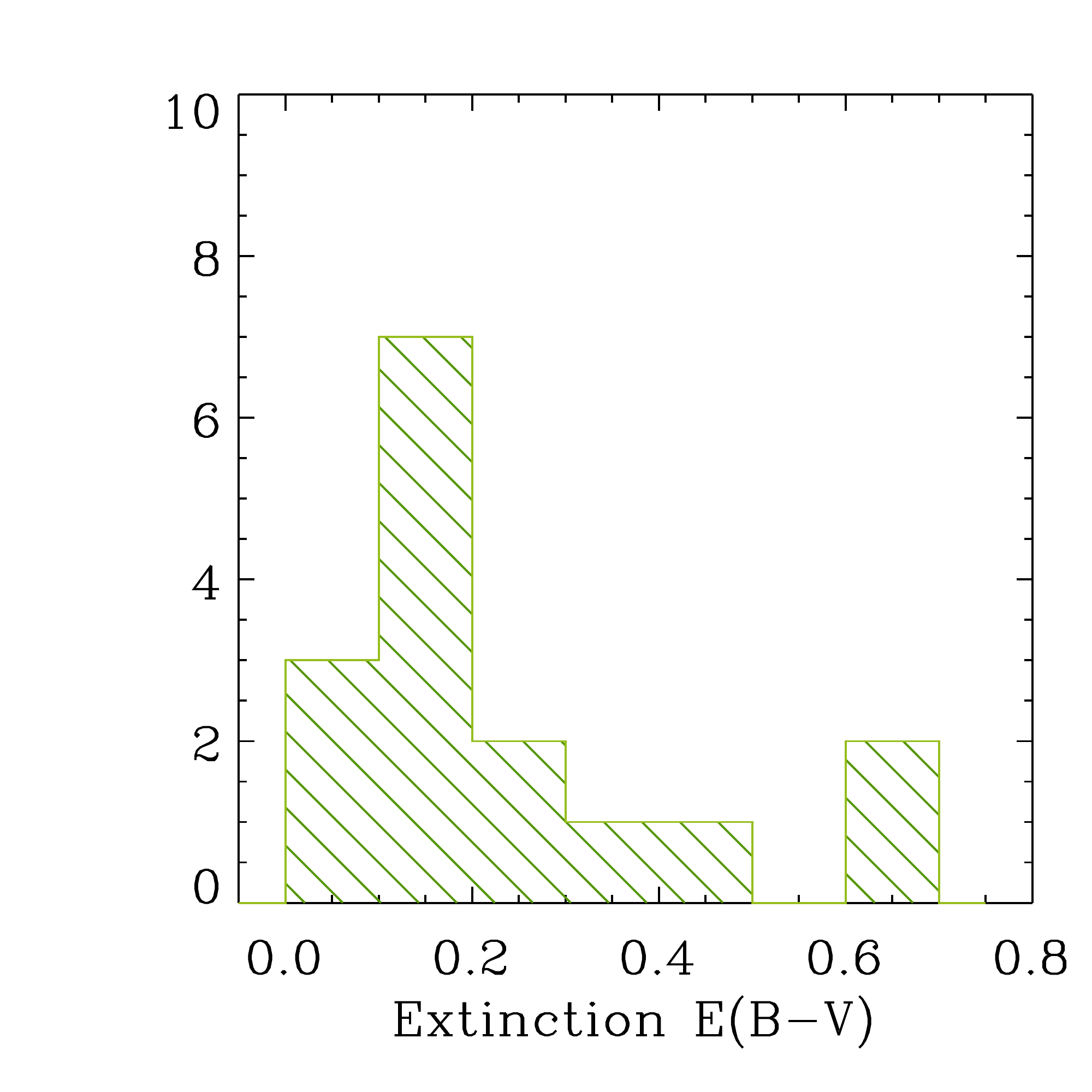}
\caption{Properties of the BBG candidates selected from the color-color criteria. From left to right: Photometric redshift distribution showing that most of the color selected BBGs has Phot-z distributions that are consistent with the redshift range of interest here. Mass distribution showing that the BBG selection is targeting the more massive galaxy population. Age distribution indicating that our selection targets evolved systems. Extinction distribution showing the range of reddening values.}
\end{figure*}

\begin{figure*}
\centering
\leavevmode
\includegraphics[scale=0.35]{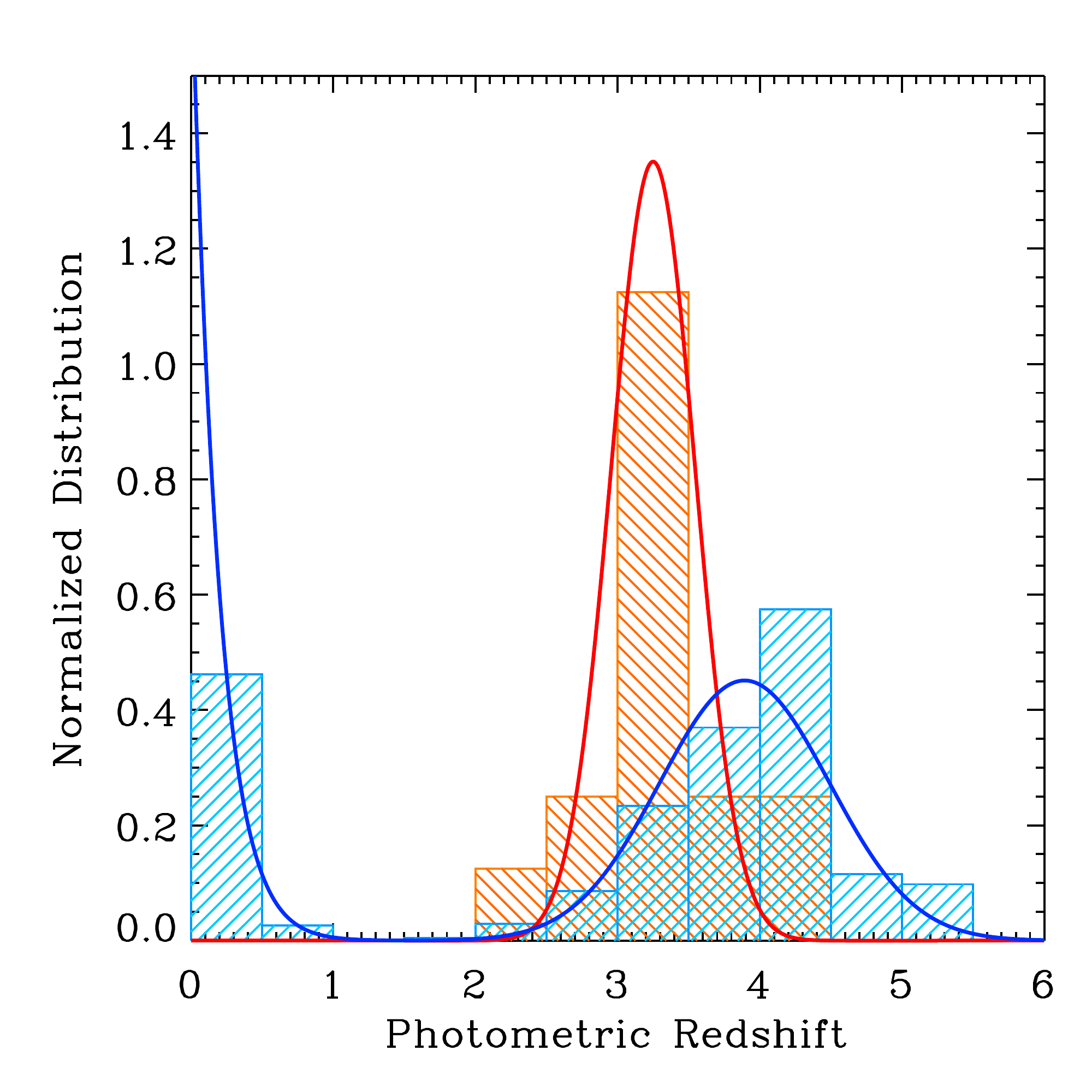}
\includegraphics[scale=0.35]{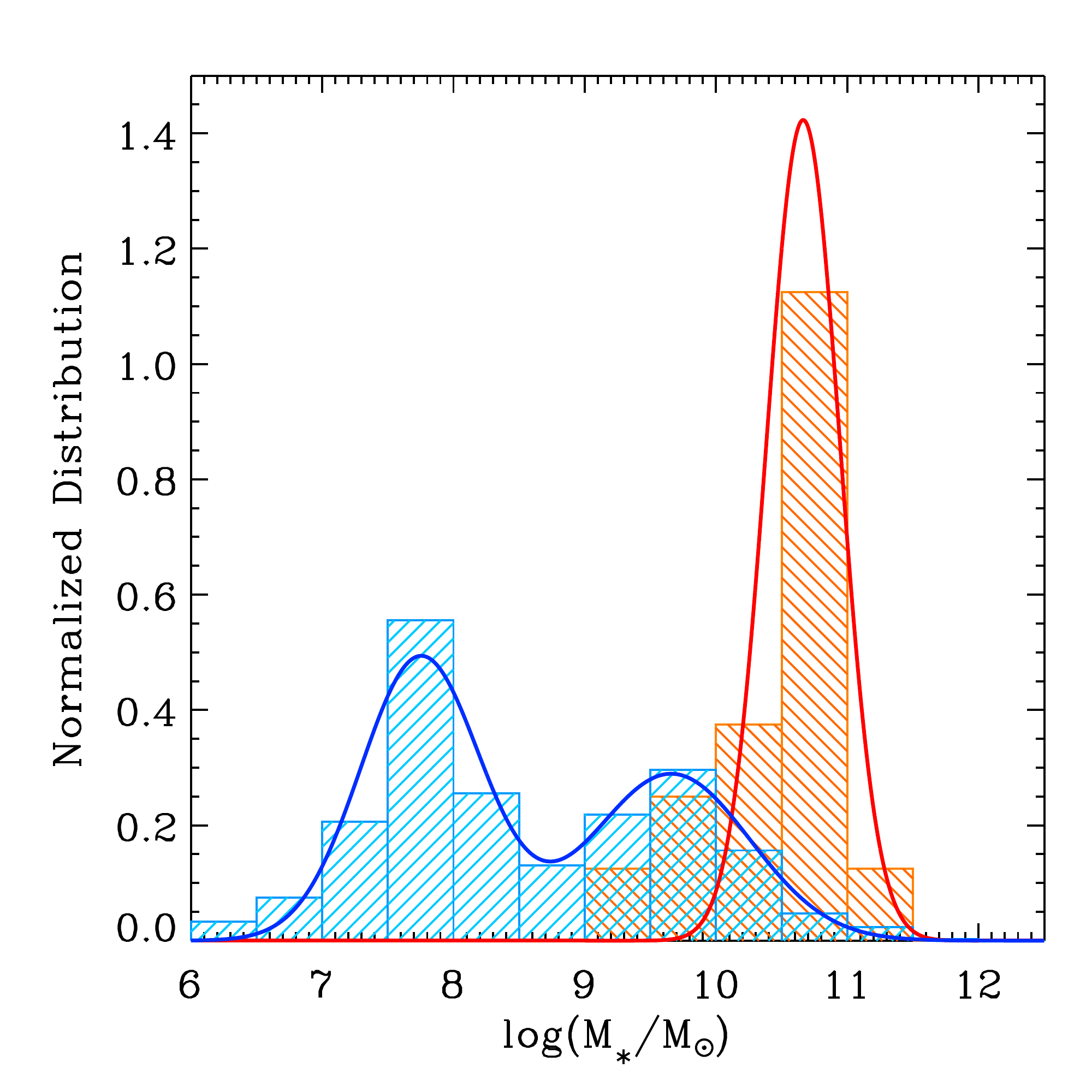} \\
\includegraphics[scale=0.35]{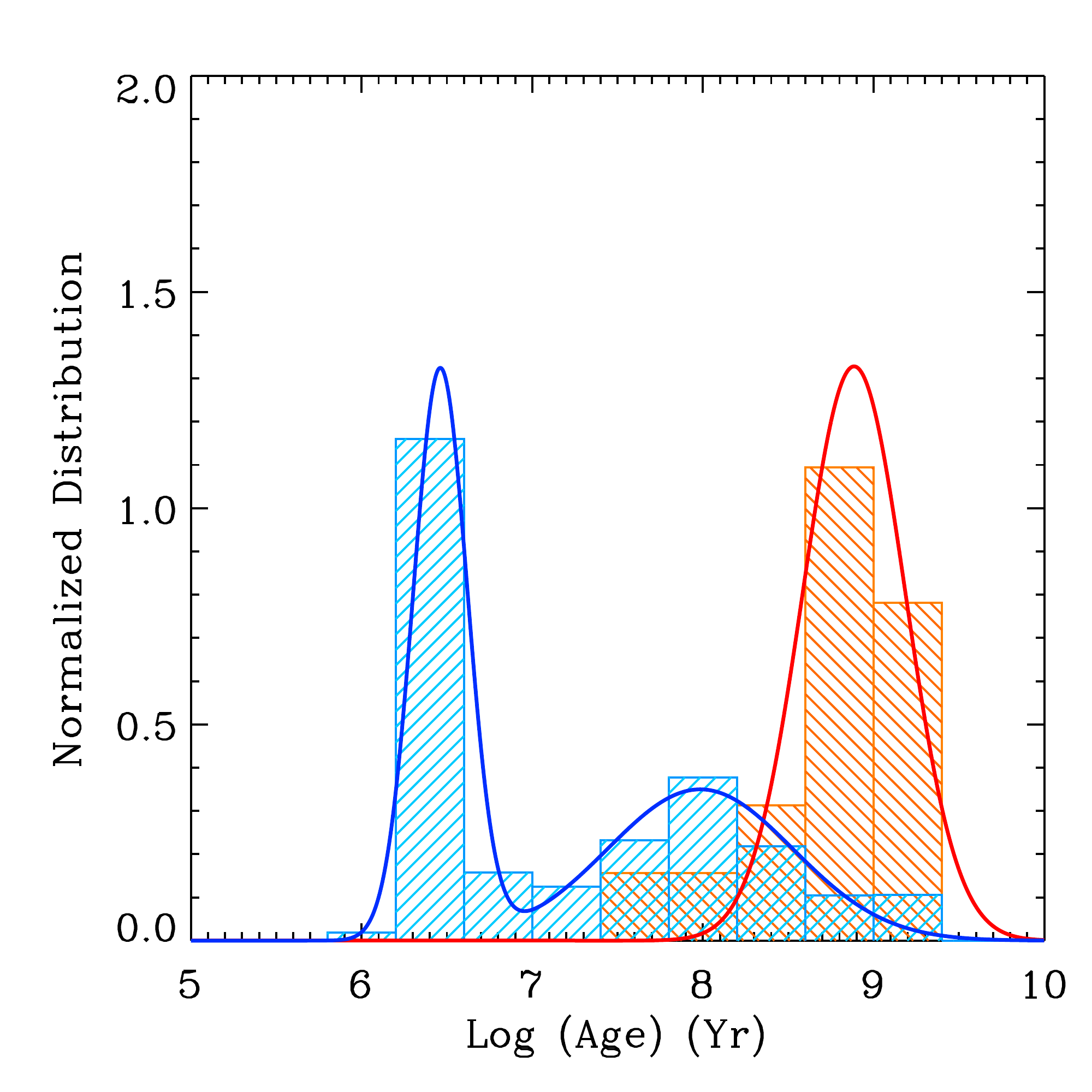}
\includegraphics[scale=0.35]{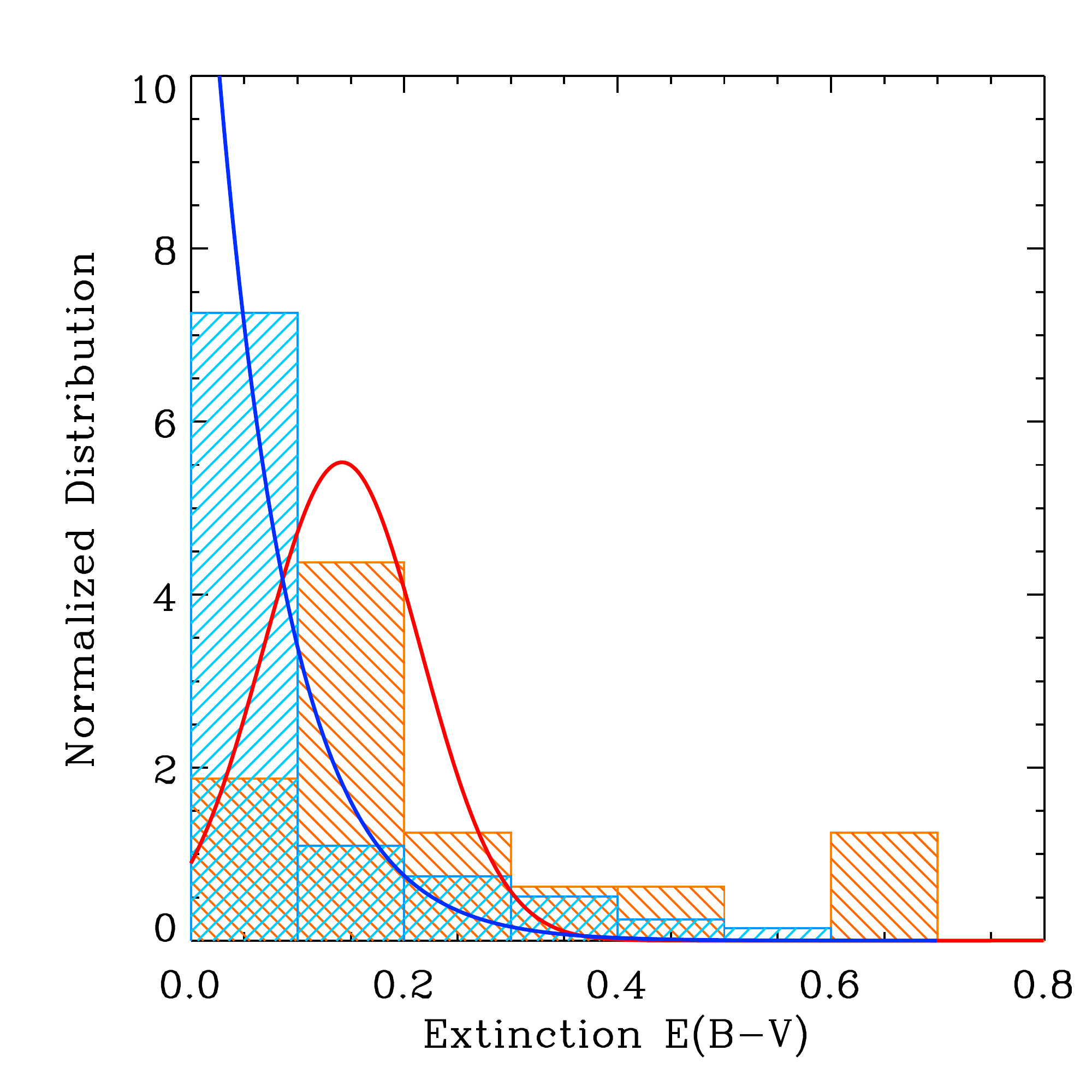}
\caption{Distribution of the properties for the LBGs (blue) and BBGs (red) in the GOODS-S from the same photometric catalog. The distributions had been normalized. Both populations are selected in the same redshift range. From the distribution, we see that the BBG selection is targeting a more evolved and massive population than the LBG selection.}
\end{figure*}

\begin{figure}
\includegraphics[trim=0cm 6cm 0cm 6cm,scale=0.42]{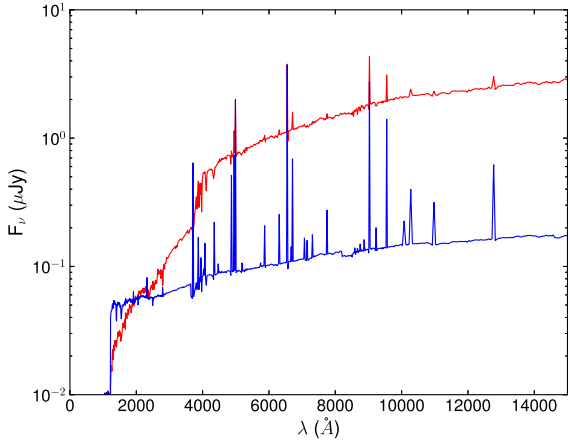} 
\caption{The stacked best-fit template SED of the BBG candidates (red) and the LBGs (blue) both normalized to the flux at 2000 \AA. The LBGs are distinguishable from BBGs by having bright rest-frame UV and strong nebular emission lines.}
\end{figure}

\begin{figure}
\includegraphics[scale=0.42]{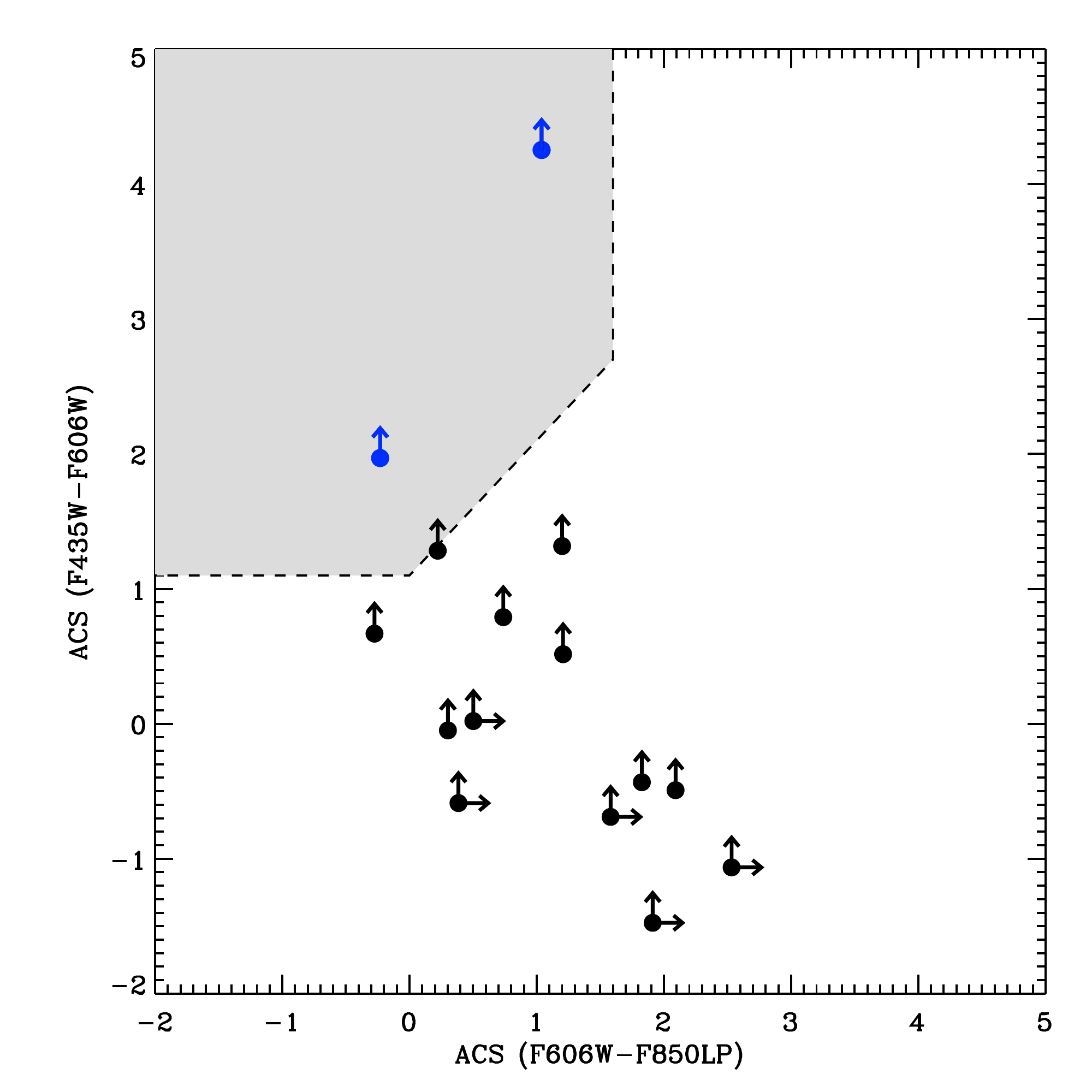} 
\caption{Optical color-color plot of LBG B drop out selection. The color selected BBG sources are plotted as black points and the color selected BBG sources that are also identified as an LBG are plotted as blue. The arrows represent $2 \sigma$ upper limits on the ACS bands for the optically faint BBGs. The B drop out selection criteria is identified as the shaded area.}
\end{figure}

We now compare photometric properties of the candidates selected by the LBG and BBG criteria. For consistency, we apply these techniques to the same parent sample of GOODS-S galaxies with the same photometry to minimize systematic effects. This approach helps us in better understanding the differences (or similarities) between the two populations of galaxies, color selected in different ways. 

The LBG selection mostly targets star forming galaxies with the pronounced Lyman break at rest-frame 912\AA\ while the Balmer break selection targets relatively quiescent systems with Balmer break feature at rest-frame 3648\AA. In principle, the BBG selection is expected to complement the LBG selection by identifying galaxies with older stellar populations, missed from the LBG selection.

In order to directly compare the BBG and LBG populations, we use the LBG selection criteria to identify galaxies over the same redshift range and area as the BBG sample in this study ($3.0<z<4.5$), using the same photometric data. This reduces sources of bias, providing a direct comparison between the two populations and selection methods. We use B-dropout selection criteria developed in \citealp {Stark2009} as the criteria for LBG selection. Using the same photometric catalog and library of model galaxies to generate template SEDs, we fitted the SED of the LBG candidates and estimated their Phot-z, stellar mass, age and extinction. Figure 12 compares the distribution of these values for the BBG and LBG populations. The BBG selected galaxies have average mass of $5 \times 10^{10} M_{\odot}$ and average age of $\sim$ 800 Myr (from the Gaussian fits to the distribution). The LBG selected galaxies on the other hand are less massive and have much younger ages. Kolmogorov$-$Smirnov test confirms that the LBG and BBG selected galaxies are drawn from statistically different populations with high confidence. This shows that the BBG selection criteria target more massive and evolved systems than the LBG selection. Using the SED inferred photometric redshifts to shift SEDs to the rest-frame, we stacked the best-fit SEDs of the BBG and LBG selected sources. Figure 13 shows the stacked rest-frame best fitted template SED of the BBG and LBG sources. The LBG selected star forming systems are bright in the rest-frame UV and are characterized by the prominent Lyman break and strong emission lines while the BBG selected quiescing systems are relatively faint in the UV and are distinguished by the break at $\sim 4000 \AA$ as we see from Figure 13. 

Figure 14 shows the distribution of the BBGs on the color-color diagram used to identify LBGs (from \citealp {Stark2009}). Also shown is the LBG locus as the grey region. Two of our color selected BBGs also satisfy LBG selection criteria (the blue filled circles in Figure 14). The two BBG selected sources that are also selected as an LBG have high estimated star formation rates and extinction suggesting that these are the dusty star-forming galaxies and likely contaminants. One of these sources (ID \#9286) has a MIPS counterpart indicative of a dusty galaxy and the other (ID \#18694) has an X-ray source associated with it, which could be an AGN. The overlap with the LBG selection is mainly due to the fact that post-starburst galaxies are in the phase of passive evolution but they still can have residual star formation present, resulting in the dual selection. The quiescent galaxies are relatively faint in the rest-frame UV (observed optical U and B bands in the redshift of interest here) as their stellar population is dominated by cooler and older stars. The BBG candidates were selected to be faint (non-detected) in the VLT/VIMOS U and HST/ACS F435W (Section 3). Candidates that have less than $2 \sigma$ detections in the optical bands are shown with upper limit arrows in Figure 14.

\section{Discussion}

Two of the widely used methods for identifying high redshift galaxies are narrow-band selection technique where line emitting galaxies at given redshift slices are selected and the LBG selection which primarily selects UV bright objects. Both techniques select actively star-forming systems and are biased against old and quiescent galaxies at high redshifts. The BBG selection technique is developed to allow for selection of these galaxies, which are missed by other techniques. Study of the nature of the BBGs would constrain star formation efficiency and mass assembly at high redshifts.

In this study we developed the criteria to identify BBGs at $3 < z < 4.5$. Several recent studies \citep {Bell2004,Brammer2011, Ilbert2013} have confirmed that the red sequence had already been in place by $z \sim 1$, that there is a significant fraction of high mass quiescent galaxies at $z \sim 2-3$ \citep {Kriek2008, Brammer2011, Marchesini2010, Whitaker2010} and support the presence of passively evolving galaxies at $z \sim 3-4$ \citep {Muzzin2013, Stefanon2013}. Furthermore, using medium band observations from ZFOURGE\footnote {http://zfourge.tamu.edu}, \citealp {Straatman2014} and \citealp {Spitler2014} identified populations of optically faint, massive and quiescent galaxies at $z \sim 4$. These studies show that although there are relatively fewer of these systems compared to the star forming galaxies at the same redshift, they still make significant contribution to the global stellar mass density of the Universe.

The most massive BBGs $(>10^{10} M_{\odot})$ in our sample have an observed number density of $\sim 3.2 \times 10^{-5}\:Mpc^{-3}$ corresponding to a stellar mass density of $\sim 2.0 \times 10^{6}\:M_{\odot}/Mpc^3$ at $3.0<z<4.5$ compared to the BBGs with no MIPS 24 micron detection which show an observed number density and mass density of $\sim 2.5 \times 10^{-5}\:Mpc^{-3}$ and $10^6\:M_{\odot}/Mpc^3$ respectively, consistent with ZFOURGE study. The BBG selection is specifically developed to minimize the contamination from dusty star forming galaxies. As discussed in Section 3.1, all the UVJ selected sources at $z \sim 4$ by \citealp {Straatman2014} fall in the BBG selection criteria although not all of them are selected as BBG candidates because of the additional non-detection condition in the observed U and $B_{435}$ bands (which probe the rest-frame UV at these redshifts). On the other hand most of the BBG selected sources at $z \sim 3-4$ fall close to the dusty star forming and quiescent boundary condition in the UVJ diagram further confirming the complications involved in distinguishing the two populations on the UVJ plane with broad band photometry at $z>3$.

\begin{table*}[t]
\begin{center}
\centering
\caption{List of the color selected BBG candidates in the CANDELS GOODS-S.}
\begin{tabular}{*{21}{c}}
\hline
\hline
ID \footnote{CANDELS GOODS-S TFIT WFC3 $H_{160}$ band selected catalog \citep{Guo2013}.}& RA & DEC & Phot-Z & log(Mass) & E(B-V) & log(Age) & SFR & F160W & (H-$K_s$) & MIPS & X-ray \\
 &  &  &  & ($M_{\odot}$) &  & (Gyr) & ($M_{\odot}Yr^{-1}$) &  &  &  & \\
\hline
2782 & 53.0835712 & -27.8875292 & 3.28 & 10.8 & 0.19 &  8.70 &  0.8 &  24.9 &  1.47 & N & N & \\
4356 & 53.1465968 & -27.8709872 & 3.88 & 11.1 & 0.69 &  8.86 &  261.2 &  26.3 &  1.56 & Y & Y & \\
4624 & 53.1379933 & -27.8682475 & 3.10 & 10.7 & 0.29 &  8.71 &  5.6 &  25.4 &  1.36 & N & Y & \\
6189 & 53.2057581 & -27.8521496 & 4.00 & 9.72 & 0.00 &  8.26 &  0.0 &  26.3 &  1.33 & N & N & \\
7526 & 53.0786782 & -27.8395462 & 4.48 & 10.6 & 0.04 &  8.71 &  0.5 &  25.9 &  1.55 & N & N & \\
9177 & 53.0383219 & -27.8252681 & 2.42 & 10.4 & 0.69 &  8.71 &  75.8 &  25.2 &  1.27 & N & N & \\
9286 & 53.2243550 & -27.8243357 & 3.27 & 10.7 & 0.39 &  9.23 &  31.8 &  25.6 &  1.24 & Y & N & \\
10479 & 53.1974865 & -27.8140435 & 3.91 & 10.5 & 0.19 &  9.16 &  9.1 &  26.4 &  1.45 & Y & N & \\
12178 & 53.0392836 & -27.7993085 & 3.22 & 10.6 & 0.04 &  9.16 &  1.8 &  25.1 &  1.28 & N & N & \\
12360 & 53.2194525 & -27.7979791 & 2.93 & 10.1 & 0.14 &  9.16 &  0.5 &  26.4 &  1.19 & N & N & \\
13327 & 53.1304511 & -27.7911228 & 3.38 & 10.1 & 0.19 &  8.26 &  2.7 &  25.6 &  1.18 & N & Y & \\
16671 & 53.1901828 & -27.7691400 & 2.69 & 10.8 & 0.19 &  9.36 &  2.1 &  24.8 &  1.20 & N & N & \\
17749 & 53.1968943 & -27.7604529 & 3.37 & 10.9 & 0.24 &  8.71 &  0.1 &  25.2 &  1.78 & N & N & \\
18180 & 53.1812226 & -27.7564225 & 3.32 & 10.8 & 0.19 &  8.71 &  0.0 &  25.1 &  1.60 & N & N & \\
18694 & 53.0503542 & -27.7521307 & 3.30 & 9.97 & 0.44 &  7.59 &  129.7 &  24.9 &  1.37 & N & Y & \\
19195 & 53.0643105 & -27.7477684 & 3.20 & 9.37 & 0.19 &  7.96 &  4.8 &  26.0 &  1.78 & N & N & \\
\hline

\end{tabular}
\end{center}
\end{table*}

The high redshift star forming galaxies are mostly selected through the LBG selection \citep {Madau1996, Steidel1999, Steidel2003, Stark2009} and the narrow band selection targeting the Ly-$\alpha$ line \citep{Kunth1998, Shapley2006, Gronwall2007}. Given the passive nature of the quiescent systems, they lack the strong emission to be targeted by the narrow band selection and they lack the strong UV continuum to be targeted by LBG selection. Using rest-frame UV luminosity function \citep {Reddy2008, Reddy2009, Bouwens2007, Bouwens2014}, we estimated an observed number density of $\sim 4-5 \times 10^{-3}\:Mpc^{-3}$ for the star forming galaxies at $<z> \sim 3-4$ (down to $L_{lim}=0.04\:L^{\star}_{z=3}$), indicating that the relatively passive galaxies at $z \sim 3-4$ (identified from BBG selections) are less than 10\% as numerous as the star forming systems at similar redshifts.

The mass function of galaxies has been studied extensively over the past few years out to $z \sim 4-5$, for both the star forming and quiescent galaxies \citep {Dickinson2003, Fontana2006, Rudnick2006, Elsner2008, Perez2008, Ilbert2013, Muzzin2013}. Using a Spitzer IRAC selected sample \citealp {Perez2008} estimated the stellar mass function of galaxies and calculated the global stellar mass density of $\sim 2.7 \times 10^{7}\:M_{\odot}/Mpc^3$ at $3.0<z<3.5$. In more recent studies, \citealp {Marchesini2010} and \citealp {Muzzin2013} estimated the stellar mass function of galaxies out to $z \sim 4$, using near infrared selected sample of galaxies. These studies estimate a stellar mass density of $\sim 4 \times 10^6\:M_{\odot}/Mpc^3$ in $3.0<z<4.0$ for the most massive $(>10^{10} M_{\odot})$ systems. Given the uncertainties inherent in the global stellar mass functions and corresponding mass densities at these redshifts, the BBG selected passive candidates could contribute as much as 10-40\% to the stellar mass density of the Universe at $z \sim 3-4$. 

With an average age of $\sim$ 800 Myr, the BBGs are expected to have formed the bulk of their mass in a burst of star formation at much higher redshifts and have been living a relatively quiescent life ever since ($z \sim 6$; \citealp {Mobasher2005, Wiklind2008, Straatman2014, Toft2014}). Assuming a constant star formation history, these massive galaxies must have assembled their mass with star formation rates $\sim 50-100\:M_{\odot}yr^{-1}$. With a bursty star formation history, which is more favored for these systems as supported by the strength of the Balmer break and the passive evolution, the progenitors of these systems must be forming stars well in excess of $100\:M_{\odot}yr^{-1}$. Recent works indicate young and highly star forming Lyman-$\alpha$ emitting systems at $z \sim 6-7$ identified through narrow band detections with star formation rates as high as few hundred $M_{\odot}/yr$ \citep {Shimasaku2006, Ouchi2010, Mallery2012, Ono2012}. Although these systems could potentially be the progenitors of the quenching systems that we identify at $z \sim 4$, their predicted number density is very small \citep {Straatman2014}. Recent studies have indicated the presence of very highly star forming galaxies at high redshift identified as dusty star forming galaxies detected in the far infrared and sub-millimeter wavelengths \citep {Younger2007, Younger2009, Magnelli2012, Karim2013}. These sub-millimeter galaxies (SMGs) are among the most star forming systems with SFRs $\gsim 1000\:M_{\odot}Yr^{-1}$ \citep {Capak2008, Fu2013}. The intense star formation is believed to be the result of gas rich major mergers \citep {Tacconi2008, Ivison2012, Fu2013}. Based on a 870 $\mu m$ selected sample of sub-millimeter selected galaxies, \citealp {Toft2014} showed that these systems could be the progenitors of the quiescent galaxies found at $z \sim 2$. By studying the number density of quiescent systems at $z \sim 4$, \citealp {Straatman2014} showed that the high redshift SMGs might have the right number density to be the progenitors of these systems. Given the similar number densities for the BBGs in the same redshift range and assuming a relatively short formation time scale for these systems ($\sim$ 200 Myr, as supported by the bursty models; \citealp {Straatman2014}) we see that the BBG selected passive systems could be the descendants of the intensely star forming sub-millimeter galaxies at $z \gsim 5-6$.

At $z \sim 3$, the most massive BBG candidates ($M_{\star} > 10^{10} M_{\odot}$) will be residing in host Dark Matter Halos (DMHs) of $M_h \sim 10^{12}\:M_{\odot} (M_{\star}/M_h \sim 0.01)$ according to semi analytic models \citep {Behroozi2013}. These systems could potentially constrain the hierarchical galaxy formation scenarios (the $\Lambda$-CDM models), if their number density exceeds the DMH number density required to host them at their redshifts \citep {Wiklind2008}. The observed number density of the BBGs is approximately 2 dex smaller than the predicted DMHs at these redshifts ($\sim 3 \times 10^{-3}\:Mpc^{-3}$; \citealp {Mo2002, Behroozi2013}). This indicates that the formation of these systems is acceptable within the hierarchical scenarios of galaxy formation although the exact mechanism of rapid mass growth is still controversial. Whether these galaxies formed their mass monolithically in these very massive halos at high redshift through constant gas accretion or through sequential merging, the physical process of this assembly is still controversial. A test of this would be to search for BBGs at higher redshifts \citep {Wiklind2008}. A large number density for the BBGs at $z \sim 5-7$ favors the monolithic collapse hypothesis as galaxies at those redshifts ($\sim$ 1 Gyr after the Big Bang) have not had enough time to go through the mergers process and to consume all their gas making stars \citep {Mobasher2005, Wiklind2008}.

\section{Summary} 

\begin{itemize}

\item We select relatively quiescent galaxy candidates based on their observed near infrared colors in the HST/WFC3 filters using the age dependent Balmer break in post-starburst systems to develop the BBG selection criteria. We identify 16 candidates in the GOODS-S Deep area at $z \sim 3-4$.

\item The nebular emission lines play a significant role in estimating the SED inferred parameters of the BBG candidates as they could alter mass and age measurements of the potential candidates. 

\item There is an overlap between the BBG selection and an LBG selection in the same redshift range of study. The BBG sample which is also selected as an LBG happens to have very high SED inferred star formation rates. Also we see that the BBG selection is picking a more massive and more evolved population compared to the LBG selection.

\item The most massive ($M_{\star} > 10^{10} M_{\odot}$) BBG selected candidates have observed number densities of $3.2 \times 10^{-5}\:Mpc^{-3}$ ($2.5 \times 10^{-5}\:Mpc^{-3}$ for the MIPS non-detected sample) and observed mass densities of $2.0 \times 10^{6}\:M_{\odot}/Mpc^3$ ($10^6\:M_{\odot}/Mpc^3$ for the MIPS non-detected sample) at $z \sim 3-4$. The massive quiescent candidates are less than 10\% as numerous as star forming galaxies at similar redshifts and could contribute as much as 10\%-40\% to the mass budget of the Universe at the redshift of the survey.

\item The BBG candidates have estimated average ages of $\sim$ 800 Myr. These systems must have had a passive evolution after the initial burst of star formation, which would indicate a high redshift of formation for these systems. The quiescent massive BBG candidates at $z \sim 3-4$ could be the descendants of the very highly star forming galaxies found at higher redshifts.

\end{itemize}

\section*{Acknowledgment}

This work is based in part on observations made with the Spitzer Space Telescope, which is operated by the Jet Propulsion Laboratory, California Institute of Technology under a contract with NASA. Support for HST Programs GO-12060 is provided by NASA through a grant from the Space Telescope Science Institute, which is operated by the Association of Universities for Research in Astronomy, Inc., under NASA contract NAS5-26555. The authors also wish to thank the anonymous referee for thoroughly reading the manuscript and providing very useful comments and suggestions.

\bibliographystyle{apj}
\bibliography{master}

\end{document}